\documentclass[aps,pra,superscriptaddress,showpacs,twocolumn]{revtex4-1}
\usepackage{amsmath}
\usepackage{amssymb}
\usepackage[pdftex]{graphicx}
\usepackage{bm}
\usepackage[pdftex]{color}
\usepackage{cases}
\usepackage{ulem}
\usepackage{multirow}
\usepackage{braket}
\usepackage{mathtools}
\usepackage{comment}

\newcommand{\red}[1]{{\textcolor{red}{#1}}}

\usepackage{bm}

\def\stacksymbols #1#2#3#4{\def\theguybelow{#2}
    \def\verticalposition{\lower#3pt}
    \def\spacingwithinsymbol{\baselineskip0pt\lineskip#4pt}
    \mathrel{\mathpalette\intermediary#1}}
\def\intermediary#1#2{\verticalposition\vbox{\spacingwithinsymbol
      \everycr={}\tabskip0pt
      \halign{$\mathsurround0pt#1\hfil##\hfil$\crcr#2\crcr
               \theguybelow\crcr}}}

\usepackage[pdftex]{hyperref}

\begin{document}
\title{Bulk-edge correspondence and stability of multiple edge states of a $\mathcal{PT}$-symmetric non-Hermitian system by using non-unitary quantum walks}

\author{Makio Kawasaki}
\affiliation{Department of Applied Physics, Hokkaido University, Sapporo 060-8628, Japan}
\author{Ken Mochizuki}
\affiliation{Department of Applied Physics, Hokkaido University, Sapporo 060-8628, Japan}
\author{Norio Kawakami}
\affiliation{Department of Physics, Kyoto University, Kyoto 606-8502, Japan}
\author{Hideaki Obuse}
\affiliation{Department of Applied Physics, Hokkaido University, Sapporo 060-8628, Japan}



\begin{abstract}%
  Topological phases and the associated multiple edge states are studied for parity and time-reversal \((\mathcal{PT})\) symmetric non-Hermitian open quantum systems by constructing a non-unitary three-step quantum walk retaining
    $\mathcal{PT}$ symmetry in one dimension. We show that the non-unitary quantum walk has large
    topological numbers of the $\mathbb{Z}$ topological phase and numerically confirm that multiple edge states
    appear as expected from the bulk-edge correspondence. Therefore, the
    bulk-edge correspondence is valid in this case.

    Moreover, we study the stability of the multiple edge states against a symmetry-breaking perturbation so that the topological phase is reduced to
 $\mathbb{Z}_2$ from $\mathbb{Z}$. In this case, we find that the
 number of edge states does not become one unless a pair of edge states
 coalesce at an exceptional point. Thereby, this is a new kind of
 breakdown of the bulk-edge correspondence in non-Hermitian systems.
 The mechanism of the prolongation of edge states against
 the symmetry-breaking perturbation is unique to non-Hermitian
 systems with multiple edge states and anti-linear symmetry.
    Toward experimental verifications, we
    propose a procedure to determine the number of multiple edge states from the time evolution of the probability distribution.
\end{abstract}

\maketitle

\section{Introduction}
\label{sec:introduction}

In closed quantum systems, all observables including Hamiltonians are described by Hermitian operators.
The hermiticity of an observable ensures that all eigenvalues are real, and the time-evolution operator generated by a Hamiltonian becomes a unitary operator.
In 1998, however, Bender and Boettcher showed that non-Hermitian Hamiltonians retaining combined parity and time-reversal ($\mathcal{PT}$) symmetry can possess entirely real eigenvalues \cite{Bender_1998}.
A non-Hermitian Hamiltonian has {$\mathcal{PT}$} symmetry if the Hamiltonian is invariant under simultaneous transformation of parity inversion and time-reversal operation \cite{Bender_1998,Mostafazadeh_2002_1,Mostafazadeh_2002_2,Mostafazadeh_2002_3,Bender_2007}.
Such non-Hermitian Hamiltonians are introduced to phenomenologically describe open systems where particles flow in and out, corresponding to gain and loss effects, for instance.
Systems described by $\mathcal{PT}$-symmetric Hamiltonians are effectively realized in classical optical systems with balanced gain and loss in experiments \cite{Guo_2009,Rutter_2010,Peng_2014_1,Peng_2014_2}.
In these systems, various phenomena that do not occur in closed systems are observed; e.g., unidirectional invisibility \cite{Lin_2011,Regensburger_2012,Mostafazadeh_2013} and sensitivity
enhancement of metrology \cite{Wiersig_2014,Wiersig_2016,Liu_2016,Chen_2017,Hodaei_2017}.
Taking these advantages, the $\mathcal{PT}$-symmetric open systems have attracted great attention to realize novel devices \cite{Feng_2018,Ganainy_2018,Ozawa_2019}.

In contrast, $\mathcal{PT}$-symmetric Hamiltonians in open quantum systems had not been realized in experiments for almost two decades since the first theoretical proposal \cite{Bender_1998}.
Recently, $\mathcal{PT}$-symmetric open quantum systems have been realized in quantum optical systems by adopting postseletions \cite{Tang_2016,Xiao_2017}.
Among them, a quantum walk approach proposed theoretically in Ref.\ \cite{Mochizuki_2016} provides a systematic way to incorporate fertile non-unitary dynamics retaining $\mathcal{PT}$ symmetry and
extra properties. By using the $\mathcal{PT}$-symmetric non-unitary quantum walk, topological phases and the edge states for open quantum systems have been studied
theoretically \cite{Kim_2016} and experimentally \cite{Xiao_2017}.
The topological number is also detected by using non-unitary quantum
walks \cite{Zhang_2017,Xiao_2018}.

The topological phase for non-Hermitian Hamiltonians (including the $\mathcal{PT}$-symmetric one) has been a rapidly growing research field in the past couple of years \cite{Esaki_2011,Kim_2016,Leykam_2017,Gong_2018,Shen_2018,Kunst_2018,Yao_2018,Xiong_2018,Kawabata_2018,
Ezawa_2019,Ghatak_2019, Borgnia_2019,Kawabata_2019,Yokomizo_2019}.
In Hermitian systems, the number of edge states appearing in a band gap near the boundary of two regions
is equivalent to the difference of topological numbers in each region.
This fundamental principle is called the bulk-edge correspondence and
its validity is widely accepted. In non-Hermitian systems, however, the bulk-edge correspondence
requires further verifications for the following reasons. First, two kinds of band gaps, point and line gaps, exist for complex energy \cite{Gong_2018,Kawabata_2018,Borgnia_2019}.
Next, the 10 symmetry classes in Hermitian systems are increased to 38
in non-Hermitian systems due to extra symmetries \cite{Kawabata_2018}.
Furthermore, a breakdown of the ordinary bulk-edge correspondence has already been reported \cite{Yao_2018,Xiong_2018,Borgnia_2019}, though the reason for it is not yet fully understood \cite{Yokomizo_2019,Imura_2019}.
The bulk-edge correspondence in non-Hermitian systems should be seriously investigated for various systems since these results strongly depend on the enlarged symmetries and details of systems.

In this work, we focus on the bulk-edge correspondence of $\mathcal{PT}$-symmetric open quantum systems with large topological numbers by using a non-unitary quantum
walk. To this end, we introduce a non-unitary three-step quantum
walk with $\mathcal{PT}$ symmetry in one dimension, which can be realized in a quantum optical
system. Because of large topological numbers, this quantum walk is expected to
exhibit multiple edge states if the bulk-edge correspondence is valid
in the open quantum system.
We numerically confirm the validity of the bulk-edge
correspondence in the $\mathcal{PT}$-symmetric open system by counting
the number of eigenvalues corresponding to multiple edge states.
We also study the stability of the multiple edge states against a symmetry-breaking perturbation so that the topological phase is reduced to
$\mathbb{Z}_2$ from $\mathbb{Z}$.
We show that, despite the perturbation, the
 number of edge states does not become one unless a pair of edge states
 coalesce at an exceptional point. Therefore, multiple edge states
 survive from the perturbation breaking the $\mathbb{Z}$ topological
 phase.
 This is a breakdown of the bulk-edge
 correspondence in non-Hermitian systems, which is different from the
 previous one \cite{Yao_2018}. The mechanism of the robustness against
 the symmetry-breaking perturbation is unique to non-Hermitian
 systems with multiple edge states and anti-linear symmetry.
Towards experimental verifications showing the existence of multiple edge states,
we also propose a procedure to distinguish the number of edge states from the time dependences of the probability distribution which can be observed in the standard experiments of quantum walks.

This paper is organized as follows. We explain the $\mathcal{PT}$-symmetric non-unitary quantum walks studied in the previous work \cite{Xiao_2017} in
Sect.\ \ref{sec:pre}. In Sect.\ \ref{sec:top}, we introduce a
non-unitary three-step quantum walk with $\mathcal{PT}$ symmetry. We
show that the quantum walk also has extra symmetries, which are
important for topological phases in open quantum systems. In addition, we
show that this quantum walk has large topological numbers and
numerically check the validity of the bulk-edge correspondence.
In Sect.\ \ref{sec:stability}, we introduce a time-evolution operator
with a symmetry-breaking perturbation so that  the topological phase is reduced to
$\mathbb{Z}_2$ from $\mathbb{Z}$ and
show the new kind of breakdown of bulk-edge correspondence in
non-Hermitian systems, which is different from the previous one \cite{Yao_2018}.
Towards future experimental verifications, in Sect.\ \ref{sec:count}, we
show that the number of edge states can be determined by the time
dependence of the probability distribution obtained from the perturbed time-evolution
operator.
Section \ref{sec:summary} closes the paper by summarizing our
results.

\section{\(\mathcal{PT}\)-symmetric quantum walks}
\label{sec:pre}
We briefly review the definition of discrete time quantum walks in one dimension \cite{Kempe_2003,Lovett_2010,Kitagawa_2012,Asboth_2013,Obuse_2015}. A quantum walk is defined by a time-evolution operator \(U\). The walker's state is characterized by its position \(x\in\mathbb{Z}\) and internal degrees of freedom \(L\) and \(R\). We describe the bases of the internal space as \(\ket{L}=(1, 0)^{\mathrm{T}}\) and \(\ket{R}=(0, 1)^{\mathrm{T}}\). For a given initial state \(\ket{\psi(0)}\), the state after \(t\) time steps is described as
\begin{equation}
\ket{\psi(t)}=U^t\ket{\psi(0)}.
\end{equation}
A state of the walker at a time step \(t\) is described as
\begin{equation}
\ket{\psi(t)}=\sum_x[a_x(t)\ket{x}\otimes\ket{L}+b_x(t)\ket{x}\otimes\ket{R}],
\end{equation}
where \(a_x(t)\) and \(b_x(t)\) are amplitudes of the states \(\ket{x}\otimes\ket{L}\) and \(\ket{x}\otimes\ket{R}\), respectively.
In most cases, time-evolution operators consist of two kinds of unitary operators, the so-called coin and shift operators:
\begin{align}
C&\coloneqq\sum_x\left[\ket{x}\bra{x}\otimes \tilde{C}_x\right], \\
S&\coloneqq\sum_x\left(\ket{x-1}\bra{x}\otimes\ket{L}\bra{L}+\ket{x+1}\bra{x}\otimes\ket{R}\bra{R}\right), \label{eq:shift}
\end{align}
respectively.
\(\tilde{C}_x\in\mathrm{U(2)}\) acts on the internal states. The shift operator \(S\) moves the walker's position depending on its internal states. Usually, the time-evolution operator is also a unitary operator since the time-evolution operator of quantum walks is defined by combining these two unitary operators.

Remarkably, quantum walks can further describe non-unitary dynamics by introducing a non-unitary operator to incorporate phenomenological gain and loss of amplitudes \cite{Regensburger_2012,Xiao_2017,Mochizuki_2016,Kim_2016,Zhang_2017,Xiao_2018,Regensburger_2011,Boutari_2016}.
Note that non-unitarity of the time-evolution operator is equivalent to non-hermiticity of the effective Hamiltonian, where the effective Hamiltonian \(H_{\mathrm{eff}}\) is defined as \(U=e^{-iH_{\mathrm{eff}}}\).
The non-unitary time-evolution operator implemented in Ref.\ \cite{Xiao_2017} is described as below:
\begin{align}
&U_2=GSR[\theta_2(x)]G^{-1}SR[\theta_1(x)], \label{eq:U_2}\\
&R[\theta(x)]=\sum_x\left[\ket{x}\bra{x}\otimes\begin{pmatrix}
\cos\theta(x) & \sin\theta(x) \\
\sin\theta(x) & -\cos\theta(x)
\end{pmatrix}\right], \label{eq:coin_exp}\\
&G=\sum_x\ket{x}\bra{x}\otimes\begin{pmatrix}
e^{\gamma} & 0 \\
0 & e^{-\gamma}
\end{pmatrix}\label{eq:g}.
\end{align}
\(S\) and \(R\) represent the shift operator given in Eq.\ \eqref{eq:shift} and the
coin operator, respectively. \(G\) is a non-unitary operator that
describes a simultaneous amplification and decay process of the walker's
amplitudes. We call \(\gamma\) the degree of non-unitarity since variations
of amplitudes become larger as \(\gamma\) increases. Since \(U_2\) in
Eq.\ \eqref{eq:U_2} contains two shift operators, \(U_2\) is a
non-unitary extension of two-step quantum walks. As we will mention
later, \(U_2\) possesses \(\mathcal{PT}\) and other extra symmetries.

We define a quasi-energy \(\varepsilon\) by an eigenvalue equation:
\begin{equation}
U\ket{\phi}=\lambda\ket{\phi},~\varepsilon=i\log\lambda.
\end{equation}
Unitarity of the time-evolution operator ensures \(|\lambda|=1\), and thus $\varepsilon$ becomes real.
However, $\varepsilon$ becomes complex in the case of non-unitary time-evolution operators.
In both cases, real part of \(\varepsilon\) has \(2\pi\) periodicity.

Here we give a remark on the topological phases of the non-unitary time-evolution operator $U_2$.
Due to the symmetries that $U_2$ has, the band gap closes at \(\varepsilon=0\) and \(\pi\) and two kinds of
topological numbers \(\nu_0\) and \(\nu_{\pi}\), respectively, are
defined.
The topological numbers \(\nu_0,\nu_{\pi}\) calculated from the
time-evolution operator \(U_2\) take the values \(\pm1\) depending on the coin parameters \(\theta_1, \theta_2\) in Eq.\ \eqref{eq:U_2}.
If we assume that the bulk-edge correspondence still holds in this system, two edge states should emerge when the difference in the topological numbers is two.
However, the time-evolution operator \(U_2\) induces only a single edge state in the standard experimental setup of quantum walks for the following reasons.
First, by changing the order of the basis to distinguish even and odd sites, \(U_2\) is described by the block diagonal form:
\begin{align}
&U_2=\begin{pmatrix}
U_e^e & 0 \\
0 & U_o^o
\end{pmatrix},\label{eq:u2dash}\\
&\ket{\psi(t)}=\binom{\ket{\psi_e(t)}}{\ket{\psi_o(t)}},
\end{align}
where \(\ket{\psi_e(t)}\) and \(\ket{\psi_o(t)}\) denote states located in even and odd sites described as
\begin{align}
\ket{\psi_e(t)}&=\sum_{i\in\mathbb{Z}}[a_{2i}(t)\ket{2i}\otimes\ket{L}+b_{2i}(t)\ket{2i}\otimes\ket{R}],  \label{eq:sl1}\\
\ket{\psi_o(t)}&=\sum_{i\in\mathbb{Z}}[a_{2i+1}(t)\ket{2i+1}\otimes\ket{L}+b_{2i+1}(t)\ket{2i+1}\otimes\ket{R}]\label{eq:sl2},
\end{align}
and \(U_e^e\) and \(U_o^o\) are non-unitary operators acting on states located in even and odd sites, respectively.
Since \(U_2\) in Eq.\ \eqref{eq:u2dash} has a block diagonal form, the
system is decoupled into two independent subsystems.
Second, the initial state is chosen to localize at a single site in many experiments on quantum walks.
Accordingly, the time-evolution operator $U_2$ realized in experiments is effectively reduced to \(U_e^e\) or \(U_o^o\) in Eq.\ \eqref{eq:u2dash}.
Actually, the topological numbers \(\nu_0\) and \(\nu_{\pi}\), derived from \(U_e^e\) or \(U_o^o\), take \(\pm \frac{1}{2}\).
Therefore, only a single edge state is realized in the experiment implementing \(U_2\) \cite{Xiao_2017}. Hence, it is insufficient to experimentally verify the bulk-edge correspondence of large topological numbers for non-Hermitian systems.


\section{Non-unitary three-step quantum walks with \(\mathcal{PT}\) symmetry}
\label{sec:top}
In order to overcome the problem that we clarified at the end of the
previous section, we define a non-unitary time-evolution operator
that is able to induce
large topological numbers and investigate this topological property in
the present section. We first introduce a time-evolution operator, called a non-unitary three-step quantum walk with \(\mathcal{PT}\) symmetry in Sect.\ \ref{subsec:top}. Then, we study the eigenvalue distribution in homogeneous systems in Sect.\ \ref{subsec:eigen_dist}, and show that the present quantum walk has large topological numbers in Sect.\ \ref{subsec:topo_number}. Finally in Sect.\ \ref{subsec:bec} we numerically check the bulk-edge correspondence in the non-unitary three-step quantum walk.

\subsection{Non-unitary three-step quantum walk with \(\mathcal{PT}\) symmetry}
\label{subsec:top}
\begin{figure}[tb]
\centering
\includegraphics[width=\columnwidth]{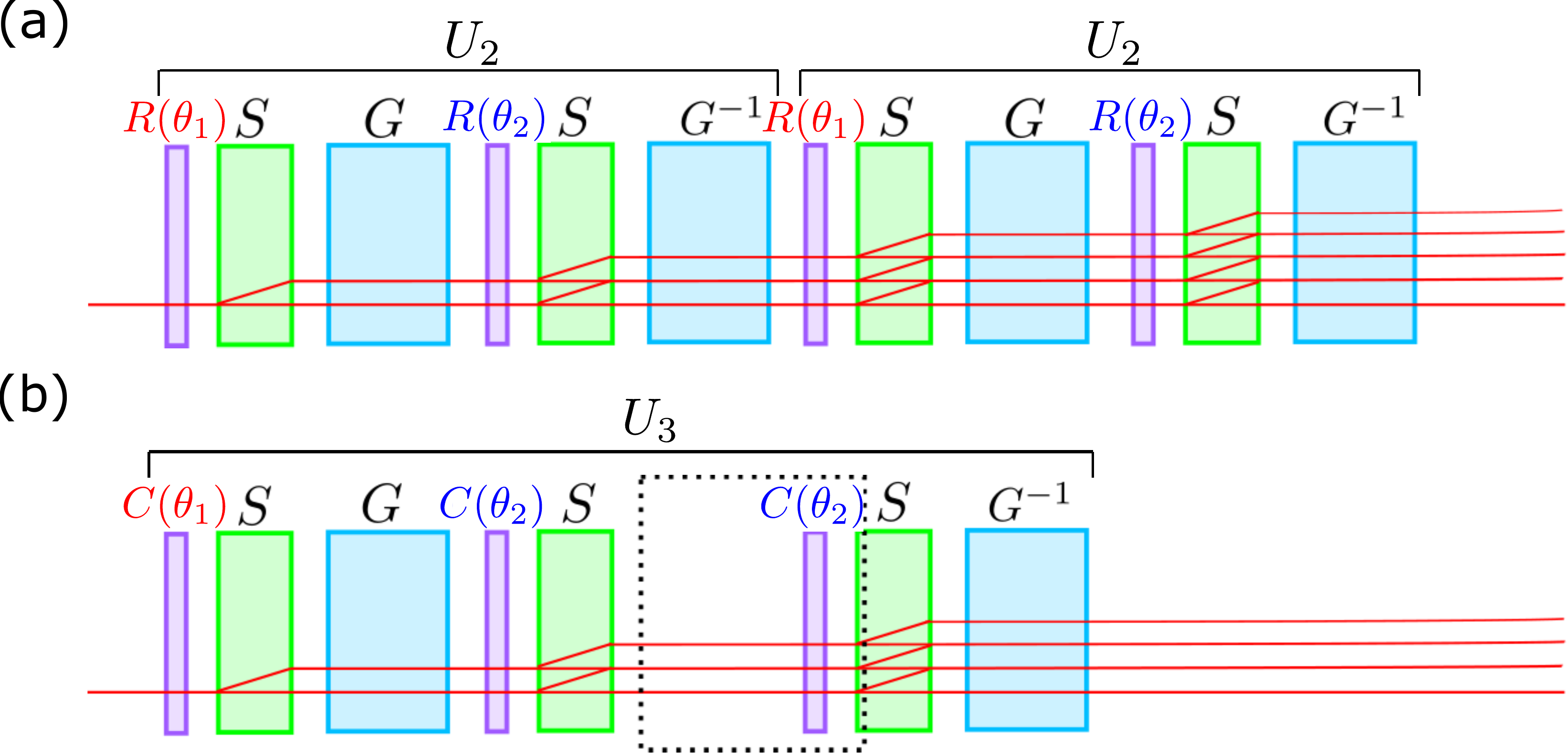}
\caption{(a) A schematic view of the experimental setup in
 Ref.\ \cite{Xiao_2017}. We use the polarization of a photon as internal
 states. The coin and shift operators and the non-unitary operator \(G\) are
 realized by half wave plates, beam displacers, and partially polarizing
 beam splitters, respectively. (b) A schematic view of the optical
 system that can realize the non-unitary three-step quantum walk \(U_3\). \(R(\theta)\) and \(C(\theta)\) are related via \(R(\theta)=C(\theta)\sigma_3\), where we denote the Pauli matrices as \(\sigma_1,\sigma_2,\sigma_3\) in this paper. We can convert \(R(\theta)\) into \(C(\theta)\) by adding an extra half wave plate acting as \(\sigma_3\).}
\label{fig:exp}
\end{figure}

We define the time-evolution operator of the non-unitary three-step quantum walk as
\begin{equation}
U_3\coloneqq G^{-1}SC[\theta_2(x)]SC[\theta_2(x)]GSC[\theta_1(x)]. \label{eq:u_3}
\end{equation}
The non-unitary operator \(G\) and the shift operator \(S\) are defined as Eqs.\ \eqref{eq:shift} and \eqref{eq:g}, and the coin operator is defined as
\begin{equation}
C[\theta(x)]\coloneqq\sum_x\left[\ket{x}\bra{x}\otimes\begin{pmatrix}
\cos\theta(x) & -\sin\theta(x) \\
\sin\theta(x) & \cos\theta(x)
\end{pmatrix}\right].\label{eq:coin}
\end{equation}
Here we call \(\theta(x)\) a coin parameter.
We also call the time-evolution operator in Eq.\ \eqref{eq:u_3} the (non-unitary) three-step quantum walk as Eq.\ \eqref{eq:u_3} contains three shift operators. Figure \ref{fig:exp} (b) shows an optical system that can realize the time-evolution operator \(U_3\).
Figure \ref{fig:exp} illustrates that experiments on the non-unitary
three-step quantum walk can be realized by rearranging the experimental
setup for \(U_2\).

Here, we clarify the symmetries that the time-evolution operator \(U_3\)
retains. First of all, we explain $\mathcal{PT}$ symmetry, which is
important for the reality of quasi-energy $\varepsilon$ \cite{Bender_1998}. The time-evolution operator \(U\) has
\(\mathcal{PT}\) symmetry if there exists  a unitary operator \(\mathcal{PT}\) such that
\begin{equation}
(\mathcal{PT})U^*(\mathcal{PT})^{-1}=U^{-1},\label{eq:pts}
\end{equation}
where the symbol \(\mathcal{P}\) means the symmetry operator that contains the effect of parity inversion \(\ket{x}\to\ket{-x}\).
We can derive this equation using the corresponding conditions
for the effective Hamiltonian, $(\mathcal{PT}) H_\text{eff}^*
(\mathcal{PT})^{-1} = H_\text{eff}$; detailed derivations are provided in Ref.\ \cite{Mochizuki_2016}. To clarify the above relation for the non-unitary three-step quantum walk \(U_3\), we redefine \(U_3\) in a symmetric time frame \cite{Asboth_2013} as
\begin{equation}
U'_3\coloneqq C\left[\frac{\theta_1(x)}{2}\right]G^{-1}SC[\theta_2(x)]SC[\theta_2(x)]GSC\left[\frac{\theta_1(x)}{2}\right]. \label{eq:u_3-STF}
\end{equation}
\(U'_3\) and \(U_3\) are related by the unitary transformation
\(U'_3=C\left[\frac{\theta_1(x)}{2}\right]U_3C\left[\frac{\theta_1(x)}{2}\right]^{\dagger}\). By
using \(U_3'\) instead of \(U_3\), we identify the symmetry operator that satisfies Eq.\ (\ref{eq:pts})
\begin{align}
 \mathcal{PT}&=\sum_x\ket{-x}\bra{x}\otimes \sigma_3,
\end{align}
where $\sigma_3$ is one of the Pauli matrices:
\begin{equation}
\sigma_1=\begin{pmatrix} 0 & 1 \\ 1 & 0\end{pmatrix}, \quad
\sigma_2=\begin{pmatrix} 0 & -i \\ i & 0\end{pmatrix}, \quad
\sigma_3=\begin{pmatrix} 1 & 0 \\ 0 & -1\end{pmatrix}.
\end{equation}
We note that the coin parameter \(\theta(x)\) must be symmetric in position space because of parity inversion:
\begin{equation}
\theta_j(-x)=\theta_j(x), \label{eq:th_c}
\end{equation}
where \(j=1,2\).

Next, we consider other symmetries by taking account of a scheme to
classify topological phases in non-Hermitian Hamiltonians proposed in
Ref.\ \cite{Kawabata_2018}.
Here, we show that $U_3^\prime$ (then, $U_3$) has all symmetries of AZ$^\dagger$
symmetry \cite{Kawabata_2018}.
In this case, time-reversal symmetry (TRS$^\dagger$),
particle-hole symmetry (PHS$^\dagger$), and chiral symmetry are
defined for a non-Hermitian Hamiltonian $H$ as
\begin{align}
 T H^T T^{-1} &= H, \label{eq:TRS_h}\\
 \Xi H^* \Xi^{-1} &= -H,\\
 \Gamma H^\dagger \Gamma^{-1} &= -H \label{eq:chiral_h},
\end{align}
respectively. We note that all symmetry operators $T$, $\Xi$,
$\Gamma$ are unitary operators. The above relations can be rewritten for
a non-unitary operator $U$ whose effective Hamiltonian satisfies Eqs.\ (\ref{eq:TRS_h})-(\ref{eq:chiral_h}) as
\begin{align}
 T U^T T^{-1} &= U,\\
 \Xi U^* \Xi^{-1} &= U, \label{eq:cond_phs}\\
 \Gamma U^\dagger \Gamma^{-1} &= U.
\end{align}
We find that $U_3^\prime$ satisfies the above relations with the
following symmetry operators
\begin{align}
 T = \sum_x \ket{x}\bra{x}\otimes \sigma_1,\\
 \Xi = \sum_x \ket{x}\bra{x}\otimes \sigma_0,\\
 \Gamma = \sum_x \ket{x}\bra{x}\otimes \sigma_1, \label{eq:chiral}
\end{align}
where \(\sigma_0=\begin{pmatrix} 1 & 0 \\ 0 & 1\end{pmatrix}\). Therefore, $U_3$ belongs to the BDI$^\dagger$ symmetry class in Ref.\ \cite{Kawabata_2018}. We, however, remind the reader that $U_3$ also retains the spatial symmetry originating from $\mathcal{PT}$ symmetry in Eq.\ (\ref{eq:pts}).


We mention that, while a similar non-unitary three-step quantum walk is studied in Ref.\
\cite{Xiao_2018}, that quantum walk has completely different
symmetries. The non-unitary quantum walk in Ref.\
\cite{Xiao_2018} does not retain $\mathcal{PT}$
symmetry, but has pseudo-hermiticity (or pseudo-unitarity).
Moreover, the quantum walk does not retain TRS$^\dagger$,
but satisifies AZ-type time-reversal symmetry:
$\tilde{T} U^* \tilde{T}^{-1} = U^{-1}$ \cite{Kawabata_2018}, where \(\tilde{T}\) is a unitary operator. Therefore,
the non-unitary three-step
quantum walk belongs to different symmetry classes to
the present quantum walk $U_3$.
Since topological phases and bulk-edge correspondence in open systems
strongly depend on symmetry classes of Hamiltonians or time-evolution operators, it is important to study the topological
phases of the present quantum walk defined by $U_3$.

\subsection{Eigenvalue distributions in the homogeneous case}
\label{subsec:eigen_dist}
Here we consider the case that coin parameters \(\theta_1(x)\) and \(\theta_2(x)\) are homogeneous in space and derive the eigenvalue distribution of the time-evolution operator \(U_3'\) \eqref{eq:u_3-STF}.
We can diagonalize the time-evolution operator \(U_3'\) in Eq.\ \eqref{eq:u_3-STF} in the wave number space by the Fourier transform because of homogeneity:
\begin{equation}
\ket{x}=\frac{1}{\sqrt{2\pi}}\int_{-\pi}^{\pi}dke^{-ikx}\ket{k}. \label{eq:ft}
\end{equation}
This procedure results in the diagonal forms of the operators in Eqs.\ \eqref{eq:shift}, \eqref{eq:g} and \eqref{eq:coin} in the wave number space:
\begin{align}
S &= \int dk\left[\ket{k}\bra{k}\otimes S_k(k)\right] ~,~ S_k(k)=\begin{pmatrix}
e^{ik} & 0 \\
0 & e^{-ik}
\end{pmatrix},\\
C(\theta_j) &= \int dk\left[\ket{k}\bra{k}\otimes C_k(\theta_j)\right] ~,~
C_k(\theta_j)=\begin{pmatrix}
\cos \theta_j & -\sin \theta_j \\
\sin \theta_j & \cos \theta_j
\end{pmatrix}, \\
G &= \int dk\left[\ket{k}\bra{k}\otimes G_k\right] ~,~ G_k=\begin{pmatrix}
e^{\gamma} & 0 \\
0 & e^{-\gamma}
\end{pmatrix},
\end{align}
where \(j=1,2\).
We rewrite the time-evolution operator \(U_3'\) in Eq.\ \eqref{eq:u_3-STF} in the wave number space as
\begin{align}
U'_3=&\int dk\left[\ket{k}\bra{k}\otimes U'_k(k)\right], \label{eq:U'_3}\\
U'_k(k)=&d_0(k)\sigma_0+d_1(k)\sigma_1+id_2(k)\sigma_2+id_3(k)\sigma_3. \label{eq:U'_k}
\end{align}
The coefficients \(d_j~(j=0,1,2,3)\) are given by
\begin{subequations}
\label{4d0}
\begin{align}
d_0(k)=&-(\cos\theta_1\sin^2\theta_2+\sin\theta_1\sin2\theta_2\cosh2\gamma)\cos k
+\cos\theta_1\cos^2\theta_2\cos3k ,\\
d_1(k)=&\sin2\theta_2\sinh2\gamma\cos k ,\\
d_2(k)=&(\sin\theta_1\sin^2\theta_2-\cos\theta_1\sin2\theta_2\cosh2\gamma)\cos k
-\sin\theta_1\cos^2\theta_2\cos3k ,\\
d_3(k)=&-\sin^2\theta_2\sin k+\cos^2\theta_2\sin3k.
\end{align}
\end{subequations}
These coefficients satisfy
\begin{equation}
d_0(k)^2-d_1(k)^2+d_2(k)^2+d_3(k)^2=1.
\end{equation}
The eigenvalues of \(U'_k(k)\) are described as
\begin{equation}
\lambda_{\pm}(k)=d_0(k)\pm i\sqrt{1-d_0(k)^2},
\end{equation}
hence eigenvalues satisfying \(|\lambda_{\pm}(k)|\ne1\) appear if there exist \(k\) such
that \(|d_0(k)|>1\). If \(|d_0(k)|=1\), the eigenvalue becomes
\(\lambda_{\pm}(k)=1\) and \(-1\), which corresponds to the
quasi-energy \(\varepsilon=0\) and \(\pi\), respectively. These
points are called exceptional points.
We also state that an eigenstate of the \(\mathcal{PT}\)-symmetric time-evolution operator \(U\) breaks \(\mathcal{PT}\) symmetry if the eigenstate of \(U\) is not the eigenstate of the operator \(\mathcal{PTK}\), where \(\mathcal{K}\) is the complex conjugation operator.
The eigenstate of \(U\) breaks \(\mathcal{PT}\) symmetry if and only if the corresponding eigenvalue \(\lambda\) satisfies \(|\lambda|\ne1\).
\begin{figure*}[tb]
\centering
\includegraphics[width=2\columnwidth]{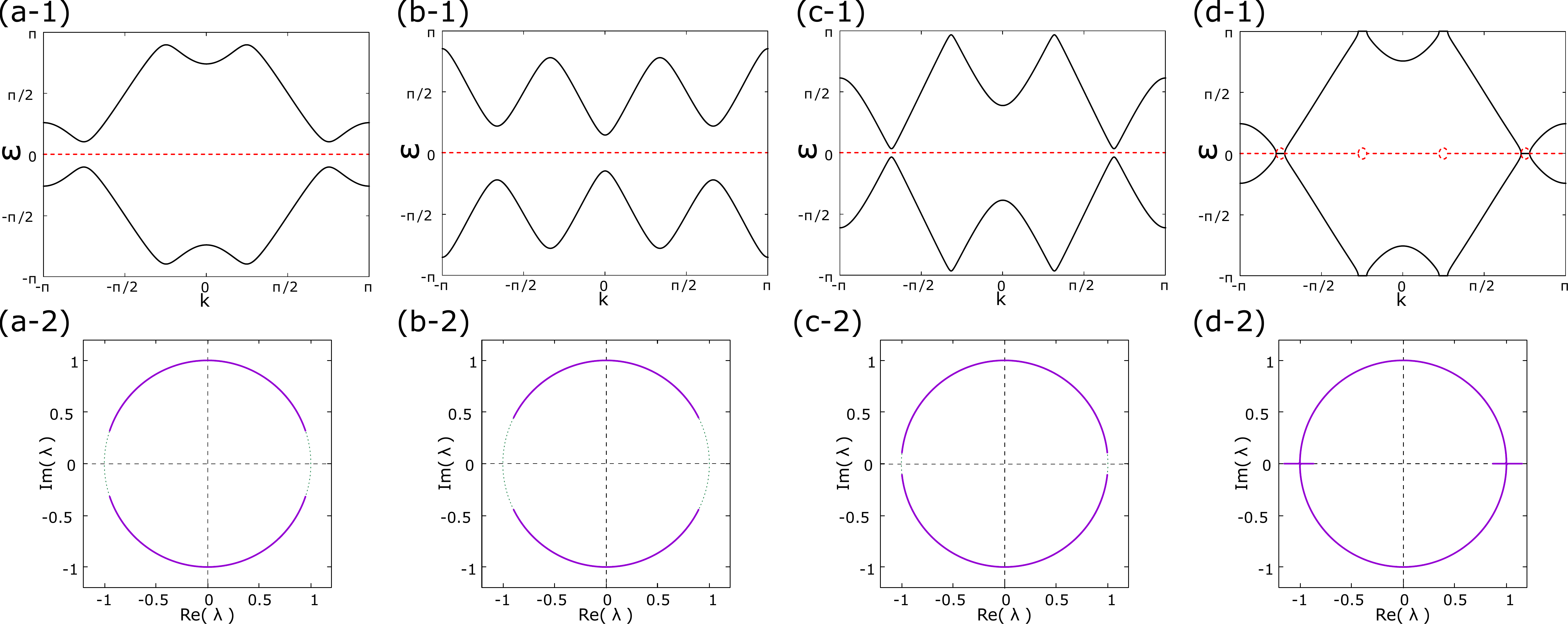}
\caption{Dispersion relations (top) and eigenvalue distributions (bottom) of the operator \(U_3'\) in Eq.\ \eqref{eq:U'_3} when \(\gamma=0.1\). Black solid (red dotted) curve represents the real (imaginary) part of \(\varepsilon\) in the top row. Green dotted and purple solid curves in the bottom row represent the unit circle of the complex plane and the eigenvalues, respectively. (a) \(\theta_1=\frac{\pi}{3}, \theta_2=\frac{\pi}{5}\). (b) \(\theta_1=-\frac{\pi}{10}, \theta_2=\frac{\pi}{8}\). (c) \(\theta_1=\frac{\pi}{10}, \theta_2=\frac{\pi}{7}\). (d) \(\theta_1=\theta_2=\frac{\pi}{4}\). }
\label{fig:dispersion}
\end{figure*}
Figure \ref{fig:dispersion} shows dispersion relations and eigenvalue
distributions of the non-unitary three-step quantum walk with
homogeneous coin parameters. In Figs.\ \ref{fig:dispersion} (a)-(c) all
eigenvalues satisfy \(|\lambda|= 1\) because none of the eigenstates
breaks \(\mathcal{PT}\) symmetry. On the other hand, in
Fig.\ \ref{fig:dispersion} (d) eigenvalues not satisfying $|\lambda|=1$ (or equivalently, complex-valued quasi-energies) emerge because several eigenstates break \(\mathcal{PT}\) symmetry.
As discussed later, the band gap closes at \(\varepsilon=0\) and \(\pi\) simultaneously.

\subsection{Topological numbers}
\label{subsec:topo_number}

Here we calculate topological numbers of the non-unitary three-step
quantum walk since the BDI$^\dagger$ class in one dimension for the real line gap has
the $\mathbb{Z}$ topological phase \cite{Kawabata_2018}.
Note that there are no constraints on values of imaginary parts of $\varepsilon$
for edge states on the real line gap.
In Hermitian systems, topological numbers of quantum walks with chiral symmetry are given by the formula \cite{Asboth_2013}
\begin{equation}
\nu_0=\frac{\nu'+\nu''}{2},\quad \nu_{\pi}=\frac{\nu'-\nu''}{2},\label{eq:top_i}
\end{equation}
where winding numbers $\nu^\prime$  and $\nu^{\prime\prime}$ are derived from the time-evolution operator with two different symmetric time frames \(U'\) and \(U''\).
We apply Eq.\ (\ref{eq:top_i}) to the present work.

Since the time-evolution operator \(U_3\) changes the positions of the walker from odd (even)
sites to even (odd) sites at every iteration (this property is called the sublattice
structure), we can rewrite the time-evolution operator $U_3$ by changing
the order of the basis, which is the same as in Eqs.\ \eqref{eq:sl1} and \eqref{eq:sl2}
\begin{equation}
\tilde{U}_3=\begin{pmatrix}
0 & U^o_e \\
U^e_o & 0
\end{pmatrix},\label{eq:u_sublattice}
\end{equation}
where \(U^o_e\) and \(U^e_o\) are non-unitary operators.
We define an operator \(\tau_3\coloneqq\begin{pmatrix}I&0 \\ 0&-I\end{pmatrix}\) in this basis, where \(I\) denotes an identity operator that has the same dimension as \(U^o_e\) and \(U^e_o\).
Then the relation
\begin{equation}
\tau_3 \tilde{U}_3\tau_3=-\tilde{U}_3, \label{eq:sublattice}
\end{equation}
holds.
This relation guarantees the existence of an eigenstate with eigenvalue
\(e^{-i(\varepsilon+\pi)}\) if and only if we have an eigenstate with
an eigenvalue \(e^{-i\varepsilon}\).
Hence the band gap closes at \(\varepsilon=0\) and \(\pi\) simultaneously and the
two kinds of topological numbers \(\nu_0\) and \(\nu_{\pi}\) take the same value.
Indeed, one of the winding numbers \(\nu''\) becomes zero.
Thus we do not distinguish the two kinds of topological numbers and denote them as \(\nu\) hereinafter.

Regarding winding numbers, we cannot calculate winding numbers in an ordinary
way because the time-evolution operator is a non-unitary operator, i.e., the effective Hamiltonian is non-Hermitian.
The method of calculating the winding number in non-Hermitian systems is presented in Ref.\ \cite{Esaki_2011} and the formula reads
\begin{equation}
\nu=\frac{1}{2\pi i}\int_{-\pi}^{\pi}dk\frac{1}{q(k)}\frac{d}{dk}q(k) .
\end{equation}
Here, \(q(k)\) is defined as
\begin{align}
\tilde{Q}(k)&=\frac{1}{2}(\ket{\chi_+}\bra{\phi_+}+\ket{\phi_+}\bra{\chi_+}-\ket{\chi_-}\bra{\phi_-}-\ket{\phi_-}\bra{\chi_-}) \nonumber\\
&=\begin{pmatrix}
0 & q(k) \\
q^*(k) & 0
\end{pmatrix}.\label{eq:Q}
\end{align}
\(\ket{\phi_{\pm}}\) and \(\bra{\chi_{\pm}}\) in Eq.\ \eqref{eq:Q} denote
right and left eigenvectors of a non-Hermitian Hamiltonian \(H(k)\) with
the eigenvalues \(E_{\pm}\), respectively, and \(H(k)\) satisfies chiral
symmetry in Eq.\ (\ref{eq:chiral_h}) with the symmetry operator
$\Gamma=\sigma_3$, thus
\begin{equation}
\sigma_3H^{\dagger}(k)\sigma_3=-H(k).\label{eq:Hcond}
\end{equation}
We apply the above method to a non-unitary time-evolution operator.
The corresponding condition for the non-unitary time-evolution operator $U(k)$ is given as
\begin{equation}
\sigma_3U^{\dagger}(k)\sigma_3=U(k).\label{eq:Ucond}
\end{equation}
In order to satisfy the condition \eqref{eq:Ucond}, we perform a unitary transformation since $U'_3$ has chiral symmetry with the symmetry operator in Eq.\ \eqref{eq:chiral}:
\begin{equation}
\begin{split}
\tilde{U}'_k(k)&=e^{-i\frac{\pi}{4}\sigma_2}U'_k(k)e^{i\frac{\pi}{4}\sigma_2}\\
&=d_0(k)\sigma_0+id_3(k)\sigma_1+id_2(k)\sigma_2-d_1(k)\sigma_3.\label{eq:U_tf}
\end{split}
\end{equation}
The eigenvectors of \(\tilde{U}'_k(k)\) in Eq.\ \eqref{eq:U_tf} are described as
\begin{align}
&\ket{\phi_{\pm}}=\frac{1}{\sqrt{2\cos2\Omega_k}}\binom{e^{\pm i\Omega_k}}{\pm ie^{\mp i\Omega_k}e^{-i\theta_k}},\\
&\bra{\chi_{\pm}}=\frac{1}{\sqrt{2\cos2\Omega_k}}(e^{\pm i\Omega_k} ,~ \mp ie^{\mp i\Omega_k}e^{i\theta_k}), \\
&d_2(k)+id_3(k)=|d(k)|e^{i\theta_k}, ~\sin2\Omega_k=\frac{d_1(k)}{|d(k)|}.
\end{align}
From the explicit form of \(q(k)\), the winding number is expressed as
\begin{equation}
\nu'=\frac{1}{2\pi}\int_{-\pi}^{\pi}\frac{d\theta_k}{dk}dk \label{eq:nu_p}.
\end{equation}
Equation \eqref{eq:nu_p} means that \(\nu'\) is equal to the winding
number of \(d_2(k)+id_3(k)\) around the origin in the complex plane. 

\begin{figure}[tb]
\centering
\includegraphics[width=\columnwidth]{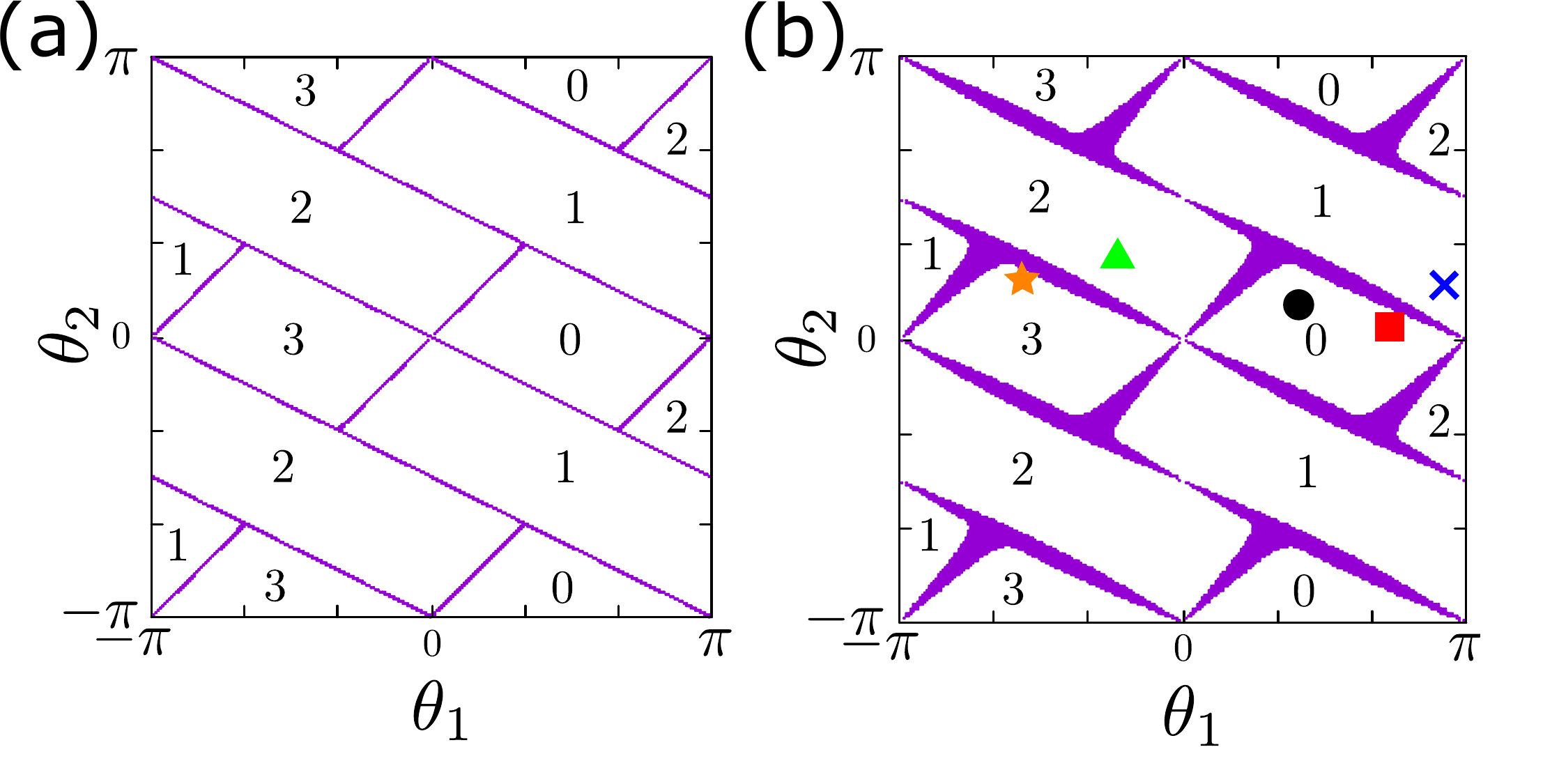}
\caption{\(\theta_1\) and \(\theta_2\) dependences of the topological
 number \(\nu\). Some bulk states break \(\mathcal{PT}\) symmetry and the band gap closes if the values \((\theta_1, \theta_2)\) are in the area shown by filled purple. Since only the difference between the topological numbers is important in the context of the bulk-edge correspondence, we add a constant \(\frac{3}{2}\) to the topological number in Eq.\ \eqref{eq:top_i} so that the minimum value becomes 0. (a) The case of unitary dynamics (\(\gamma=0\)). (b) The case of non-unitary dynamics (\(\gamma=0.1\)).}
\label{fig:topoinv}
\end{figure}

Figure \ref{fig:topoinv} shows values of the topological number
\(\nu\) as a function of the coin parameters \(\theta_1\) and \(\theta_2\).
As shown in Fig. \ref{fig:topoinv}, the values of the topological
number do not depend on the degree of non-unitarity \(\gamma\), unless the band gaps around \(\varepsilon=0\) or \(\pi\) close.
In contrast to the recent experiment \cite{Xiao_2017}, the values of the
topological number can take 2 or 3.
Such a large topological number originates from terms with the argument $3k$ of trigonometric
functions in Eq.\ (\ref{4d0}).
Actually, these terms are a consequence of the three shift operators
in the three-step quantum walk $U_3$. We can, thereby, construct quantum walks
with much larger topological numbers by increasing the number of shift operators.

\subsection{Bulk-edge correspondence}
\label{subsec:bec}
In order to verify the bulk-edge correspondence, we consider inhomogeneous systems so that the topological number varies in position space by making the coin parameters \(\theta_1\) and \(\theta_2\) position dependent. We set the parameters satisfying Eq.\ \eqref{eq:th_c} as
\begin{equation}
\theta_{1(2)}(x) = \begin{cases}
\theta_{1(2)}^{i} & (|x|<L') \\
\theta_{1(2)}^{o} & (|x|\geq L')
\end{cases},\label{eq:th_d}
\end{equation}
and thus the values of \(\theta_1\) and \(\theta_2\) change at two points \(x=\pm L'\). We call the region \(|x|<L'\) \((|x|\geq L')\) the inner (outer) region. We fix \(L'=50\) and \((\theta_1^i, \theta_2^i)=(\frac{2}{5}\pi, \frac{1}{10}\pi)\) [the black circle in Fig.\ \ref{fig:topoinv} (b)] in the rest of this section, thus the topological number of the inner region \(\nu^i\) is \(0\).  Then, we consider several sets of \(\theta_1^o\) and \(\theta_2^o\) in order to change the topological number of the outer region, \(\nu^o\).
In our cases \(\nu^o\) represents the difference between the topological numbers of the inner and outer regions due to \(\nu^i=0\).

\begin{figure*}[tb]
\centering
\includegraphics[width=1.80\columnwidth]{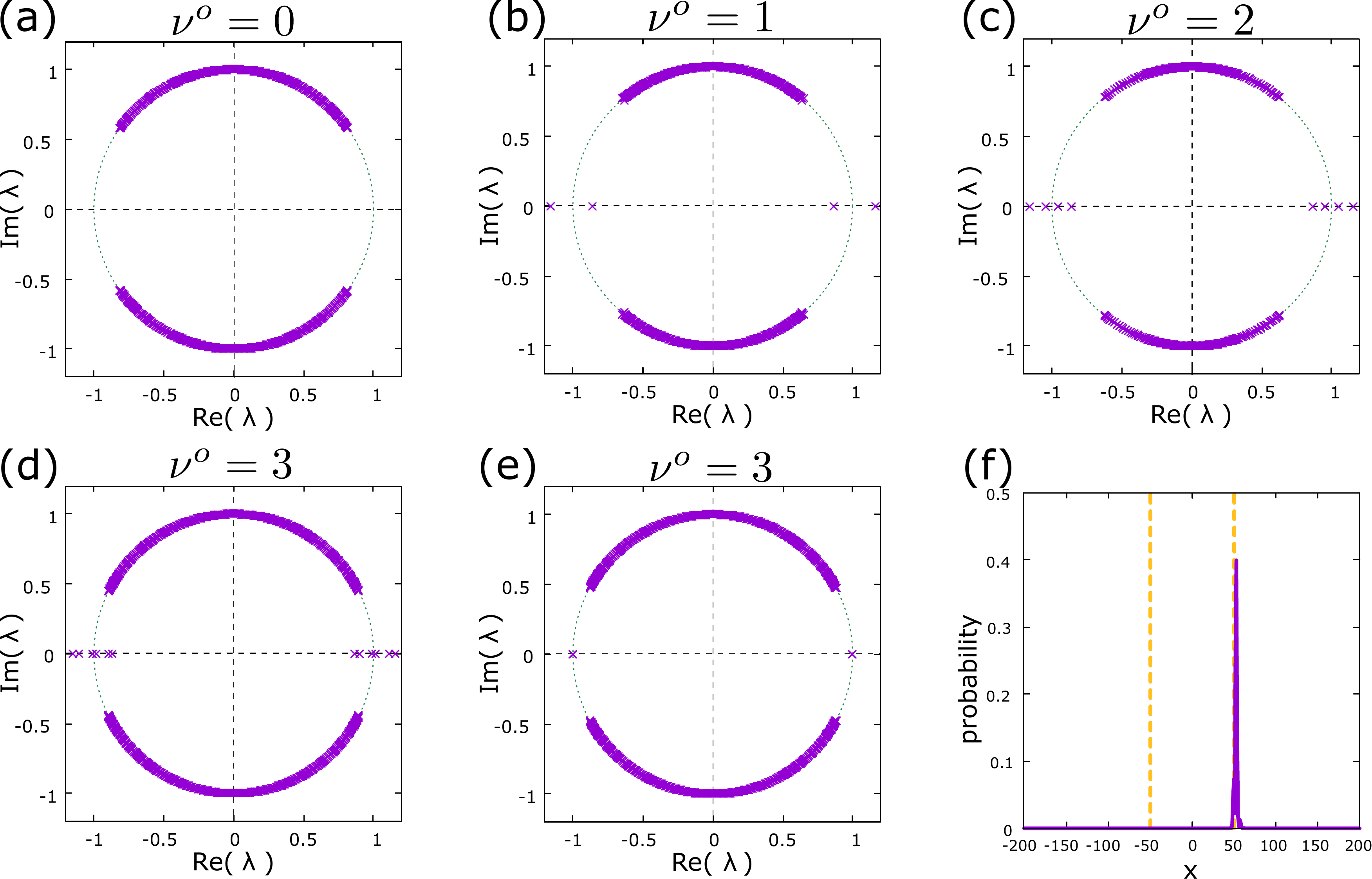}
\caption{(a)-(d) Eigenvalue distributions in the complex plane of the
 time-evolution operator \(U_3\) in the case of
 \(\gamma=0.1\). Green dotted curves represent the unit circle. \(\nu^o\)
 denotes the topological number of the outer region. (a)
 \(\theta_1^o=\frac{7}{10}\pi, \theta_2^o=\frac{1}{20}\pi\)
 (red square in Fig.\ \ref{fig:topoinv}). (b)
 \(\theta_1^o=\frac{9}{10}\pi, \theta_2^o=\frac{1}{5}\pi\) (blue
 cross). (c) \(\theta_1^o=-\frac{1}{5}\pi,\theta_2^o=\frac{3}{10}\pi\)
 (green triangle). (d)
 \(\theta_1^o=-\frac{3}{5}\pi,\theta_2^o=\frac{1}{5}\pi\) (orange
 star). (e) Eigenvalue distribution of \(U_3\) in the case of
 \(\gamma=0\). \(\theta_1^o\) and \(\theta_2^o\) are the same as in (d).
 (f) Probability distribution of the edge state whose eigenvalue
 has a minimum real part in (d). Two yellow vertical dashed lines denote the boundaries between the inner and outer regions (\(x=\pm L'\)).}
\label{fig:eigenvalue}
\end{figure*}
In Fig. \ref{fig:eigenvalue}, we plot eigenvalue distributions of the
time-evolution operator \(U_3\) in Eq.\ \eqref{eq:u_3} with inhomogeneous coin parameters.
These eigenvalues are calculated by numerical diagonalization for a finite system size (\(|x|\leq 400\)) with periodic boundary conditions.
Except \(\nu^o=0\) [Fig.\ \ref{fig:eigenvalue} (a)], isolated
eigenvalues $\lambda$
appear on the real axis.
The corresponding eigenstates are edge states of the three-step quantum walk localized near $|x|\approx L'$, whose real part of the quasi-energy \(\varepsilon\) is \(0\) or \(\pi\).
First we consider unitary time evolution [Fig.\ \ref{fig:eigenvalue} (e)].
In this case six eigenstates with eigenvalues \(\pm1\) are degenerate.
Since there are two boundaries where the topological number changes, the bulk-edge correspondence tells us that the number of edge states is twice as large as the difference in the topological numbers.
In this case, edge states with eigenvalues \(\pm1\) appear and the
number of edge states is twice as large as \(\nu^o\), which corresponds to the prediction of the bulk-edge correspondence.
Next we consider the case of non-unitary time evolution [Figs.\ \ref{fig:eigenvalue} (a)-(d)].
The eigenvalues of the edge states \(\lambda\) satisfy \(|\lambda|\ne1\) [or equivalently \(\mathrm{Im}(\varepsilon)\ne0\)] and this means that the edge states break \(\mathcal{PT}\) symmetry.
Figure \ref{fig:eigenvalue} (f) shows the probability distribution of an edge state.
This state apparently breaks \(\mathcal{PT}\) symmetry because the
probability distribution shows a peak only near the boundary \(x=L'\)
and thus is not symmetric in position space.
As shown in Figs.\ \ref{fig:eigenvalue} (a)-(d), the number of edge states
with \(\mathrm{Re}(\varepsilon)=0\) or \(\pi\) is twice as large as \(\nu^o\). Therefore, the bulk-edge correspondence holds even in a non-unitary quantum walk with large topological numbers in these parameters.

\begin{figure}[tb]
\centering
\includegraphics[width=0.75\columnwidth]{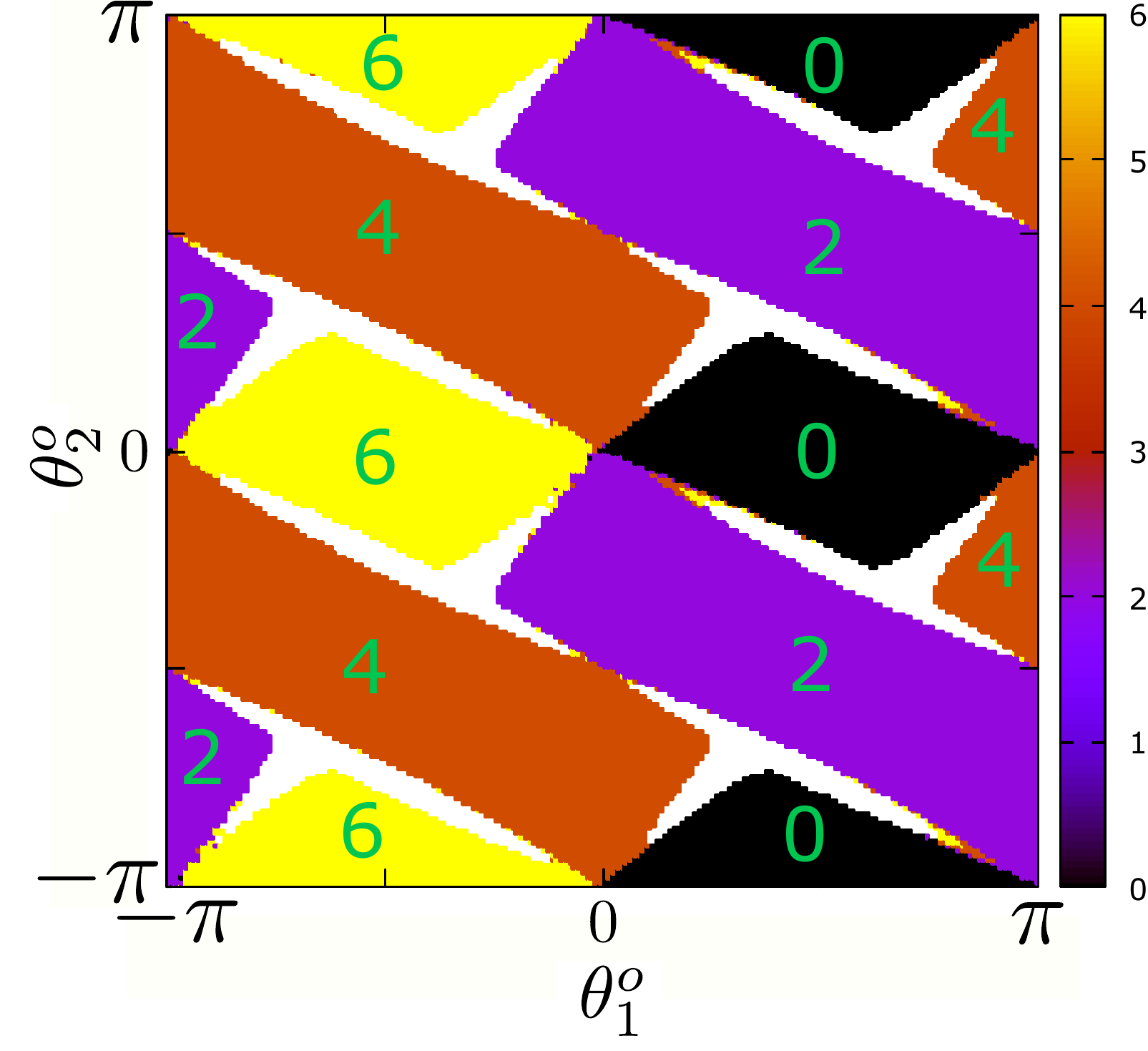}
\caption{\(\theta_1^o\) and \(\theta_2^o\) dependences of the number of
 edge states with \(\mathrm{Re}(\varepsilon)=0\). There are no edge
 states because bulk states close the band gap when \((\theta_1^o,
 \theta_2^o)\) are located in regions colored white (unfilled regions) without numbers.}
\label{fig:bec}
\end{figure}
Finally, we verify the bulk-edge correspondence in the whole parameter region of \(\theta_1^o\) and \(\theta_2^o\) for the fixed parameters \((\theta_1^i, \theta_2^i)=(\frac{2}{5}\pi, \frac{1}{10}\pi)\).
Figure \ref{fig:bec} shows the number of edge states with \(\mathrm{Re}(\varepsilon)=0\) as a function of \(\theta_1^o\) and \(\theta_2^o\).
Compared with Fig.\ \ref{fig:topoinv} (b), the number of edge states is
twice as large as the topological number of outer regions \(\nu^o\).
Hence the bulk-edge correspondence holds for the non-unitary three-step quantum walk belonging to the BDI\(^{\dagger}\) class with large topological numbers.
We also confirmed the same result for edge states with \(\mathrm{Re}(\varepsilon)=\pi\).

\section{Stability of the edge states against perturbations}
\label{sec:stability}
In this section we investigate the stability of the edge states.
To this end, we introduce a time-evolution operator with a symmetry-breaking perturbation \(U_{\delta}\),
\begin{equation}
U_{\delta}\coloneqq G^{-1}SC(\theta_2)SC(\theta_2+\delta)GSC(\theta_1),\label{eq:u_delta}
\end{equation}
where \(\delta\) denotes the strength of a perturbation.
A finite value of $\delta$ breaks time-reversal symmetry, chiral
symmetry, and \(\mathcal{PT}\) symmetry, while it keeps particle-hole
symmetry in Eq.\ (\ref{eq:cond_phs}).
Thereby, by introducing the perturbation, the symmetry class
is changed from class BDI$^\dagger$ to class D\(^\dagger\) and the topological
phases for real line gaps are reduced to $\mathbb{Z}_2$ from
$\mathbb{Z}$ for those in BDI$^{\dagger}$
\cite{Kawabata_2018}.
Note that edge states can still appear at
$\text{Re}(\varepsilon)=0$ and $\pi$ in class D$^\dagger$ and  two topological numbers $\nu_0$ and $\nu_\pi$
satisfy the relation \(\nu_0=\nu_{\pi}\) because \(U_{\delta}\)
still keeps the sublattice structures explained in Eq. (\ref{eq:sublattice}) in Sect.\
\ref{subsec:topo_number}.
We also note that particle-hole symmetry in Eq.\ (\ref{eq:cond_phs})
also ensures that the time-evolution operator \(U_{\delta}\) has complex
conjugate pairs of eigenvalues, except edge states whose
eigenvalues are real.

We obtain the topological numbers in class D$^{\dagger}$
by taking account of the reduction of topological phases from
$\mathbb{Z}$ in class BDI$^\dagger$  to $\mathbb{Z}_2$ as explained below.
For convenience we define a quantity \(\Delta\nu\) as the difference in the topological numbers of two regions before adding the perturbation
(class BDI$^\dagger$).
We naively expect that the difference in the two topological numbers of two
regions after adding the perturbation (class D$^\dagger$) is equal to $\text{mod}(\Delta
\nu,2)$ because of the reduction  from $\mathbb{Z}$ to $\mathbb{Z}_2$.

\begin{figure}[tb]
\centering
\includegraphics[width=0.90\columnwidth]{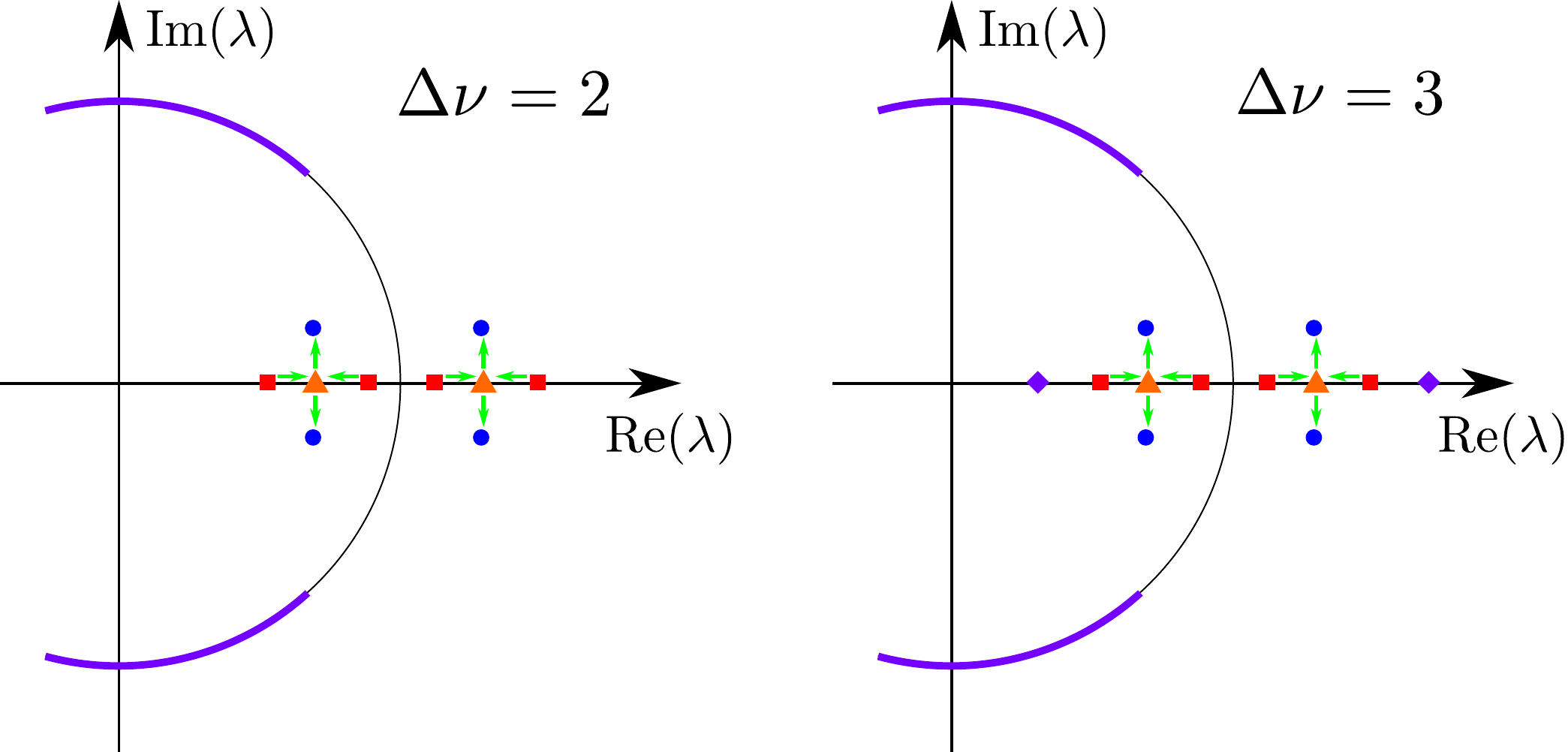}
\caption{Schematic views for the \(\delta\) dependence of
 eigenvalues of edge states of $U_\delta$ with $\Delta
 \nu=2,3$. At first, two eigenvalues corresponding to prolonged
 edge states (denoted by red squares) remain real and become close to
 each other with increasing \(\delta\). At a certain finite \(\delta\), the time-evolution operator forms exceptional points (denoted by orange triangles) and teh two eigenstates coalesce. Finally the two eigenvalues become complex, corresponding to defective states (denoted by blue circles). In the case of \(\Delta\nu=3\), two edge states have complex eigenvalues and the eigenvalue of the one remaining edge state (denoted by purple diamonds) is kept real with increasing \(\delta\).}
\label{fig:ep}
\end{figure}

Here, we consider how edge states are influenced by
the perturbation.  On one hand, when $\Delta \nu=1$, there should be no fundamental difference in the
edge state because of $\mod(\Delta \nu,2) =1$. On the other hand, the
edge states are strongly affected by the perturbation when $\Delta
\nu=2,3$. In these cases,
we expect that two edge states with real
eigenvalues $\lambda$ for \(\delta=0\) form a complex
conjugation pair with complex eigenvalues by introducing the perturbation \(\delta\ne0\), since
the $\mathbb{Z}_2$ topological phase only ensures a single edge state at most.
We refer to the pair states as {\it defective states}. While the
defective states are induced by an infinitesimally small value of
$\delta$ in unitary cases, we can predict interesting prolongations of
multiple edge states, i.e., no defective states, for a finite value of $\delta$ in non-unitary cases as
explained below. When
we regard a time-evolution operator \(U\) as a non-Hermitian Hamiltonian
\(H\), the relation of particle-hole symmetry for a time-evolution
operator in Eq.\ \eqref{eq:cond_phs} is equivalent to the relation of
time-reversal symmetry for a Hamiltonian in AZ class:
\begin{equation}
 TH^*T^{-1}=H
\end{equation}
where \(T\) is a unitary operator. In other words, the above
relation means anti-linear symmetry.
Using a similar discussion for \(\mathcal{PT}\)-symmetric non-Hermitian
Hamiltonians, we can predict that eigenvalues of edge states must be
kept real unless an exceptional point of degeneracy occurs when the
value of $\delta$ increases from zero continuously.
In unitary cases with \(\Delta\nu=2,3\), eigenvalues of the edge states
are already degenerate when \(\delta=0\) and the edge states
become defective states for any finite value of \(\delta\).
In non-unitary cases with \(\Delta\nu=2,3\), however, the eigenvalues of
edge states are not degenerate when \(\delta=0\) as shown in Figs.\
\ref{fig:eigenvalue} (b)-(d). Therefore, up to a certain value of
$\delta$, eigenvalues of all edge states remain real as
schematically explained in Fig.\ \ref{fig:ep}.
This prolongation of real eigenvalues results in
the existence of multiple edge states even in class D$^\dagger$ whose
topological phase is $\mathbb{Z}_2$, and we call such edge states {\it prolonged edge states}. (Note that the definition of
the edge states is localized states with $\text{Re}(\varepsilon)=0$ or
$\pi$.) Therefore, this is a new kind of breakdown of
bulk-edge correspondence in non-Hermitian systems.
The breakdown originates from the exceptional point for edge states due to the anti-linear symmetry of the time-evolution
operator.

\begin{figure*}[tbh]
\centering
\includegraphics[width=2\columnwidth]{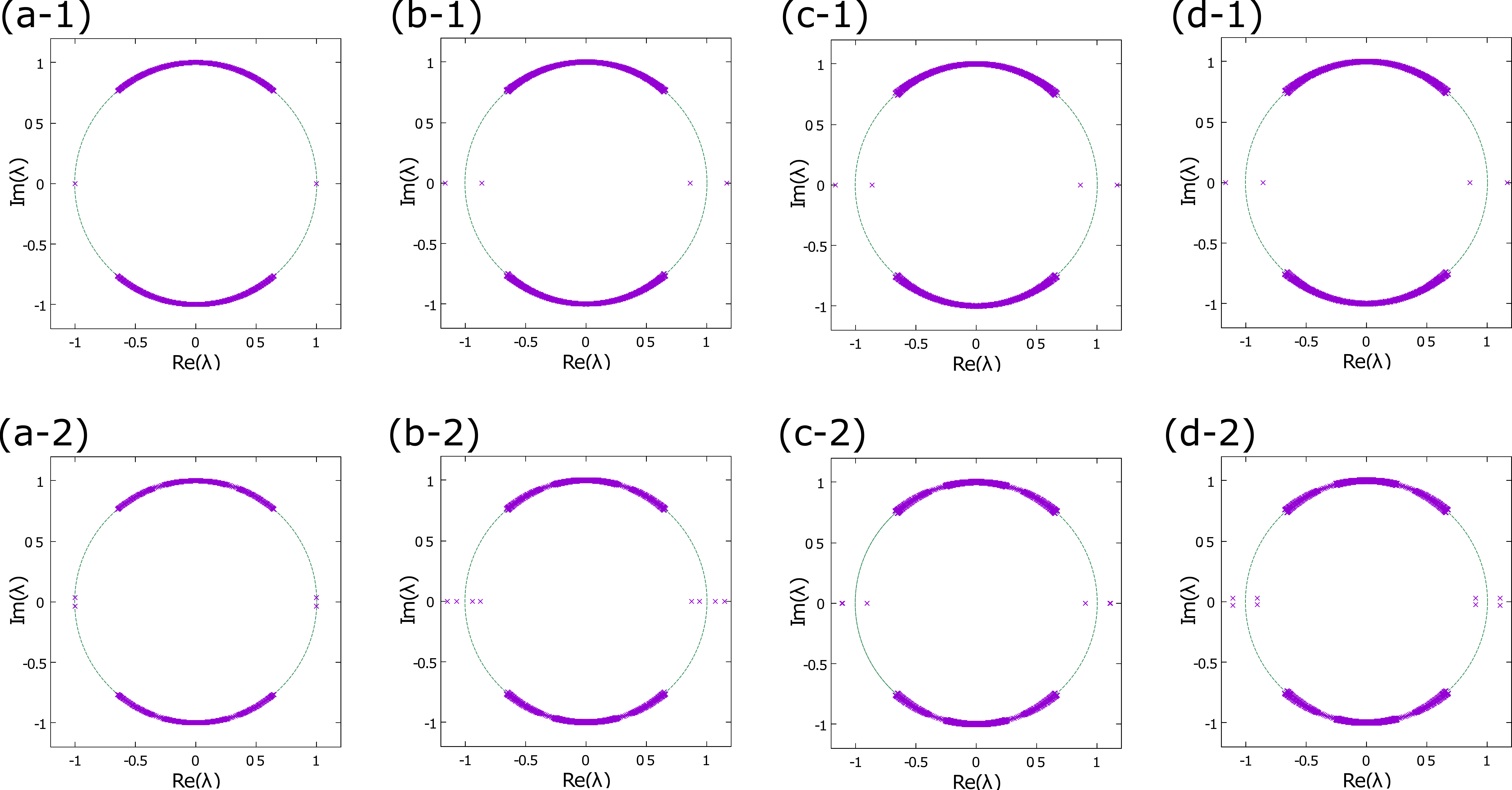}
\caption{Eigenvalue distributions of the perturbed time-evolution operator
 \(U_{\delta}\). We set
 \(\theta_1^i=\frac{2}{5}\pi,\theta_2^i=\frac{1}{10}\pi\) (\(\nu^i=0\))
 in both cases. (Top row)
 \(\theta_1^o=\frac{9}{10}\pi,\theta_2^o=\frac{1}{5}\pi\)
 (\(\nu^o=1\)). (Bottom row)
 \(\theta_1^i=-\frac{1}{5}\pi,\theta_2^i=\frac{3}{10}\pi\)
 (\(\nu^o=2\)). We also set the degree of non-unitarity \(\gamma=0.1\) except for (a). (a)  \(\delta=0.05\) and \(\gamma=0\) (unitary dynamics). (b) \(\delta=0.05\). (c) \(\delta=0.0696\). (d) \(\delta=0.08\).}
\label{fig:exceptional_point}
\end{figure*}

Here, we numerically verify the above theoretical prediction
by calculating eigenvalue distributions of the perturbed time-evolution
operator \(U_{\delta}\) with $\Delta \nu=1,2$ and show the results in Fig.\ \ref{fig:exceptional_point}.
As shown in the top panels of Fig.\ \ref{fig:exceptional_point}, the
eigenvalues of the edge states remain real in the case of \(\Delta\nu=1\)
because of $\mod(\Delta \nu,2)=1$ as expected.
In the case of $\Delta \nu=2$, however, while the eigenvalues of the edge states for \(\gamma=0\)
become complex [Fig.\ \ref{fig:exceptional_point} (a-2)] at the
small value of \(\delta=0.05\), those for \(\gamma\ne0\) remain
real [Fig.\ \ref{fig:exceptional_point} (b-2)] at the same value of
$\delta$.
With increasing \(\delta\), each pair of eigenvalues becomes degenerate, forming the exceptional point [Fig.\ \ref{fig:exceptional_point} (c-2)],
and then the pair of eigenvalues becomes a complex conjugation pair, i.e., the pair of edge states become defective states [Fig.\ \ref{fig:exceptional_point} (d-2)].
Therefore, we confirm the theoretical prediction.

\begin{figure*}[tb]
\centering
\includegraphics[width=1.80\columnwidth]{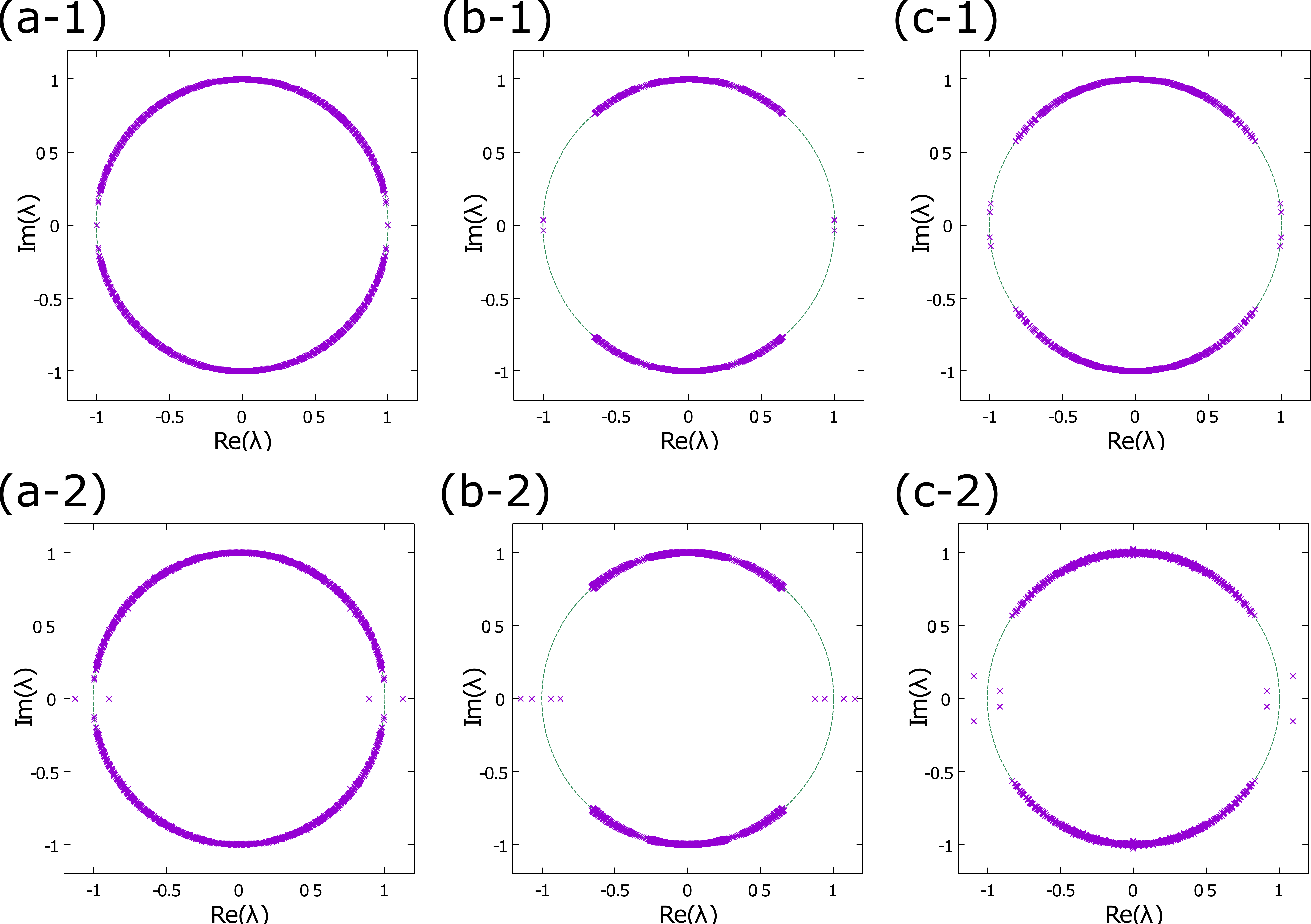}
\caption{Eigenvalue distributions of the disordered time-evolution
 operator $\tilde{U}_\delta$. We set
 \(\theta_1^i=\frac{2}{5}\pi,\theta_2^i=\frac{1}{10}\pi\) and \(\delta=0.05\) in all
 calculations. (Top row) \(\gamma=0\). (Bottom row)
 \(\gamma=0.1\). (a)
 \(\theta_1^o=\frac{9}{10}\pi,\theta_2^o=\frac{1}{5}\pi\) (\(\Delta\nu=1\))
 and \(\theta_r=0.1\). (b) \(\theta_1^i=-\frac{1}{5}\pi,
 \theta_2^i=\frac{3}{10}\pi\) (\(\Delta\nu=2\)) and \(\theta_r=0.001\). (c)
 Coin parameters are same as (b) and \(\theta_r=0.1\).}
\label{fig:ep_random}
\end{figure*}

Finally, we numerically study the stability of edge states against the disorder of
coin parameters and show that prolonged edge states are robust against
disorder.
For each coin operator, the disordered coin parameter
\(\tilde{\theta}(x)\) is given as
\(\tilde{\theta}(x)=\theta(x)+\delta\theta(x)\), where \(\theta(x)\) is
given by Eq.\ (\ref{eq:th_d}) while \(\delta\theta(x)\) is a
random number in the range \([-\theta_r,\theta_r]\) distributed
uniformly over position space. Using \(\tilde{\theta}(x)\), the
time-evolution operator is expressed as
\(\tilde{U}_{\delta}= G^{-1}SC[\tilde{\theta}_2(x)]SC[\tilde{\theta}_2(x)+\delta]GSC[\tilde{\theta}_1(x)]\).
Note that the time-evolution operator $\tilde{U}_\delta$ has particle-hole symmetry even in the presence of disorder.
Figure \ref{fig:ep_random} shows eigenvalue distributions of the disordered time-evolution operator \(\tilde{U}_{\delta}\) when \(\Delta\nu=1,2\).
As shown in Fig.\ \ref{fig:ep_random} (a), the eigenvalues of the edge
states remain real in the presence of disorder when \(\Delta\nu=1\)
for both unitary and non-unitary cases. Therefore, as expected,
the edge states are robust against disorder in the case of $\Delta \nu=1$.
Next, we study the case of $\Delta \nu=2$.
In unitary cases, defective states are immediately induced by $\delta$
and are unstable against disorder as shown in Figs.\ \ref{fig:ep_random}
(b-1) and (c-1).
In non-unitary cases, however, eigenvalues of the prolonged edge states remain
real for weak disorder [Fig.\ \ref{fig:ep_random} (b-2)],
while they become complex for stronger disorder [Fig.\ \ref{fig:ep_random} (c-2)]].
This clearly shows that prolonged edge states are not accidental
states with real eigenvalues, but rather are robust against disorder as if
ordinary edge states are topologically protected against local perturbations.

 We also confirmed the consistent results for \(\Delta\nu=3\).
In such cases, one edge state is stable against perturbations and
disorder, and the remaining two edge states become defective states
for large $\delta$ or strong disorder. Thereby,
the defective states of the non-unitary quantum walk with $\Delta
\nu=3$ are also more robust than those of the unitary quantum walk, similar to the case of \(\Delta\nu=2\).

We can apply the same discussion for a non-Hermitian Hamiltonian
belonging to class BDI$^\dagger$.
 Suppose that the Hamiltonian possesses multiple edge states whose real part of the eigenenergy
 is zero, originating from the $\mathbb{Z}$ topological phase for a real line gap in one dimension.
By introducing a perturbation only preserving PHS\(^{\dagger}\),
\begin{equation}
  \Xi H^* \Xi^{-1} = -H,
\end{equation}
the topological phase is reduced to $\mathbb{Z}_2$ in the class D$^\dagger$.
Denoting the Hamiltonian \(H\) as \(H=iH'\), we can rewrite the
relation for PHS\(^{\dagger}\) in anti-linear symmetry for \(H'\):
\begin{equation}
  \Xi H'^* \Xi^{-1}=H'.
\end{equation}
Then, the eigenvalues of edge states of $H$
must be kept pure imaginary unless they form exceptional points when we increase the strength of the perturbation.
Therefore, the number of edge states can be larger than one even in
the $\mathbb{Z}_2$ topological phase, which results in the breakdown of
the bulk-edge correspondence.

\section{Counting the number of edge states by symmetry-breaking perturbation}
\label{sec:count}
We have confirmed that the bulk-edge correspondence holds for the non-unitary three-step quantum walk by counting the number of eigenstates satisfying \(\mathrm{Re}(\varepsilon)=0\) or \(\mathrm{Re}(\varepsilon)=\pi\) in the Sect.\ \ref{sec:top}.
However, in the standard experiments on quantum walks, only the probability distribution of walkers can be observed.
\begin{figure}[tb]
\centering
\includegraphics[width=0.9\columnwidth]{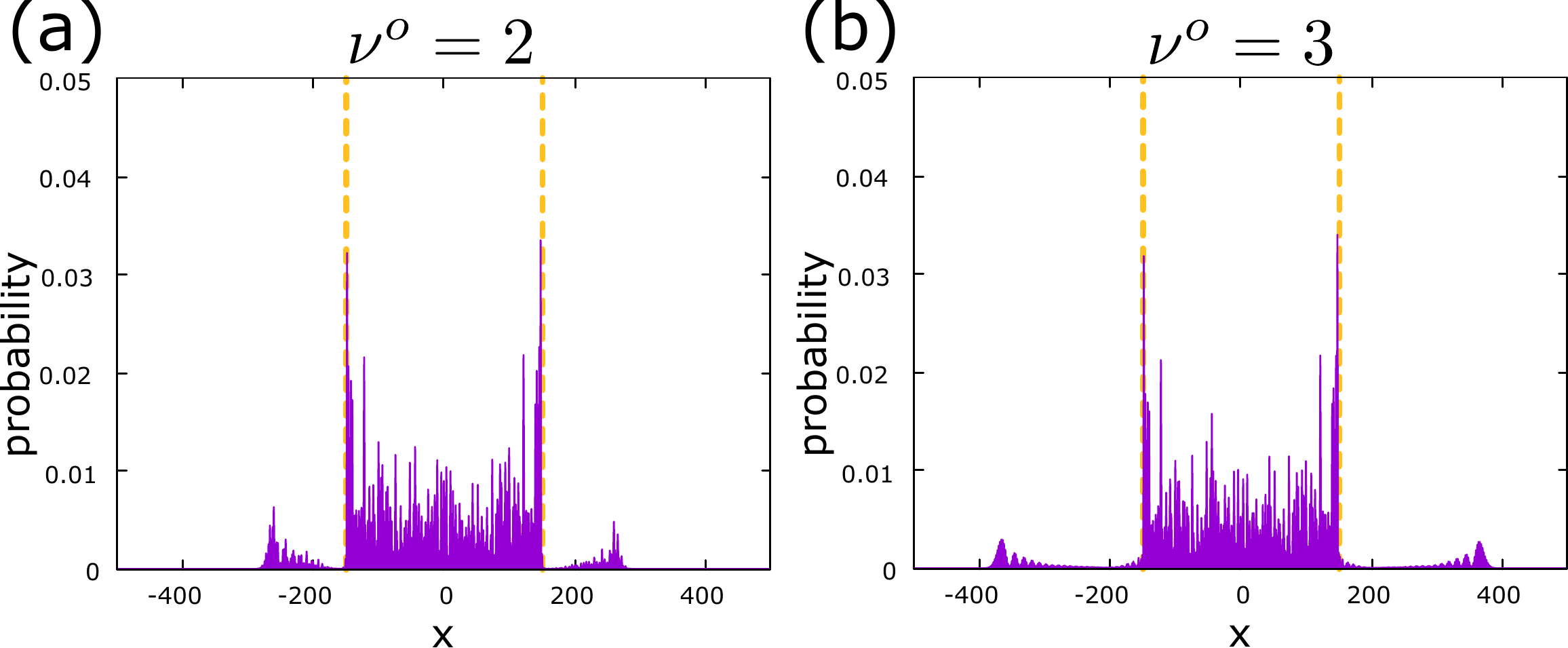}
\caption{Probability distributions of the non-unitary three-step quantum
 walk. In both cases we fix the parameters
 \(\theta_1^i=\frac{2}{5}\pi,~\theta_2^i=\frac{1}{10}\pi\), and
 \(\gamma=0.1\). The initial state and time step are given as \(\ket{\psi(0)}=\ket{0}\otimes \frac{1}{\sqrt{2}}(\ket{L}+i\ket{R})\) and 246, respectively. Yellow dashed lines represent the positions of the boundary of the coin parameters (\(x=\pm L'\)). (a) \(\theta_1^o=-\frac{1}{5}\pi, \theta_2^o=\frac{3}{10}\pi\) (\(\nu^o=2\)). (b) \(\theta_1^o=-\frac{3}{5}\pi, \theta_2^o=\frac{3}{20}\pi\) (\(\nu^o=3\)).}
\label{fig:prob_dist}
\end{figure}
Figures \ref{fig:prob_dist} (a) and (b) show probability distributions in the cases in which the difference between the topological numbers in the inner and outer regions is 2 and 3, respectively.
Apparently, we cannot determine the number of edge states from these probability distributions.
In this section, we show that we can determine the number of edge states
from the time-step dependences of probability distributions under the perturbation introduced in the previous section.

In this section we put \(\gamma=0\) to realize unitary dynamics since the value of the topological number does not depend on \(\gamma\) as long as the band gaps are open.
In this section we define the coin parameters as
\begin{equation}
\theta_{1(2)}(x) = \begin{cases}
\theta_{1(2)}^{L} & (x\leq0) \\
\theta_{1(2)}^{R} & (x>0)
\end{cases}.\label{eq:th_lr}
\end{equation}
Here, we divide the system into two regions at \(x=0\) since parity is not essential for unitary time evolution.
The quasi-energies of the defective states are shifted by \(\pm \omega_{\delta}\) because of the particle-hole symmetry.
Here, we focus on the probability of detecting the walker at \(x=0\), \(p_0(t)\).
If a pair of defective states emerges, they must give rise to interference between two defective states and \(p_0(t)\) oscillates in time.
Detailed analysis of \(p_0(t)\) is provided in Appendix \ref{app:count}.
As shown in Appendix \ref{app:count}, the size of the band gaps affects the localizing length of defective states and the behavior of \(p_0(t)\).
To characterize the size of the band gaps, we define a minimum quasi-energy for bulk states \(\varepsilon_m\) as
\begin{equation}
  \varepsilon_m\coloneqq \min\varepsilon_{\mathrm{b}\uparrow},
\end{equation}
where \(\varepsilon_{\mathrm{b}\uparrow}\) denotes a quasi-energy of a upper-band bulk state.
\begin{figure}[tb]
\centering
\includegraphics[width=0.5\columnwidth]{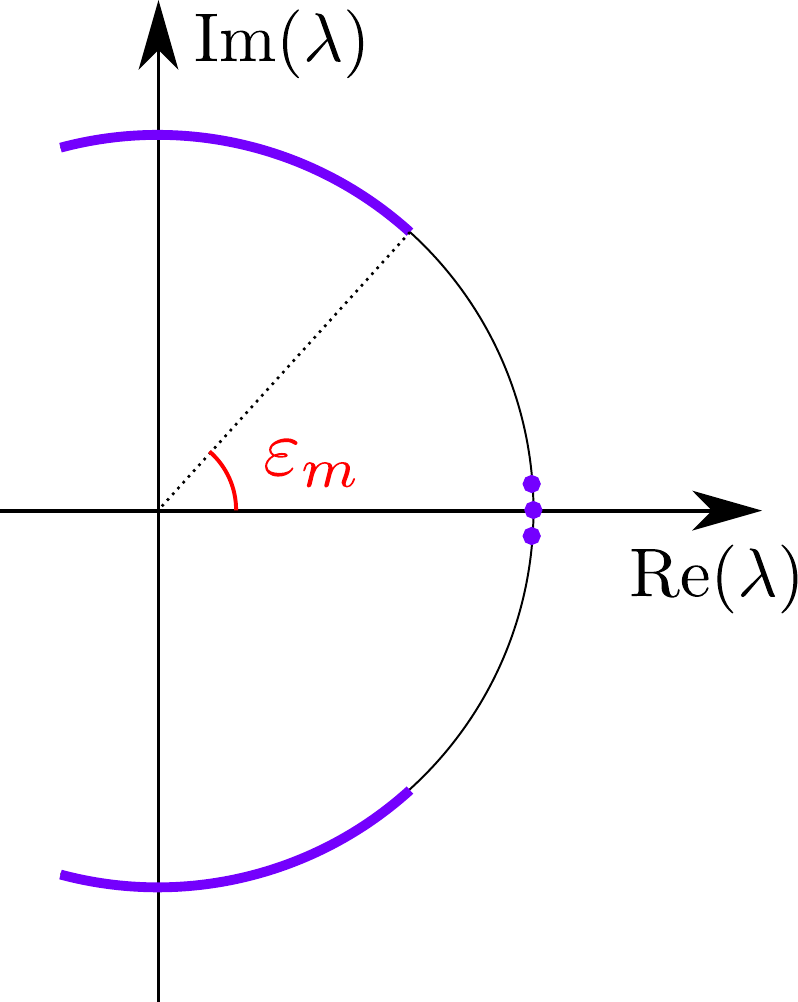}
\caption{Schematic of a minimum quasi-energy for bulk states \(\varepsilon_m\). The black circle represents a unit circle. Purple lines and dots represent the eigenvalues of the time-evolution operator for bulk and edge states, respectively.}
\label{fig:bandgap}
\end{figure}
The schematic meaning of \(\varepsilon_m\) is provided in Fig.\ \ref{fig:bandgap}.
The size of the band gaps is large if the value of \(\varepsilon_m\) is large.

\begin{figure*}[tb]
\centering
\includegraphics[width=2\columnwidth]{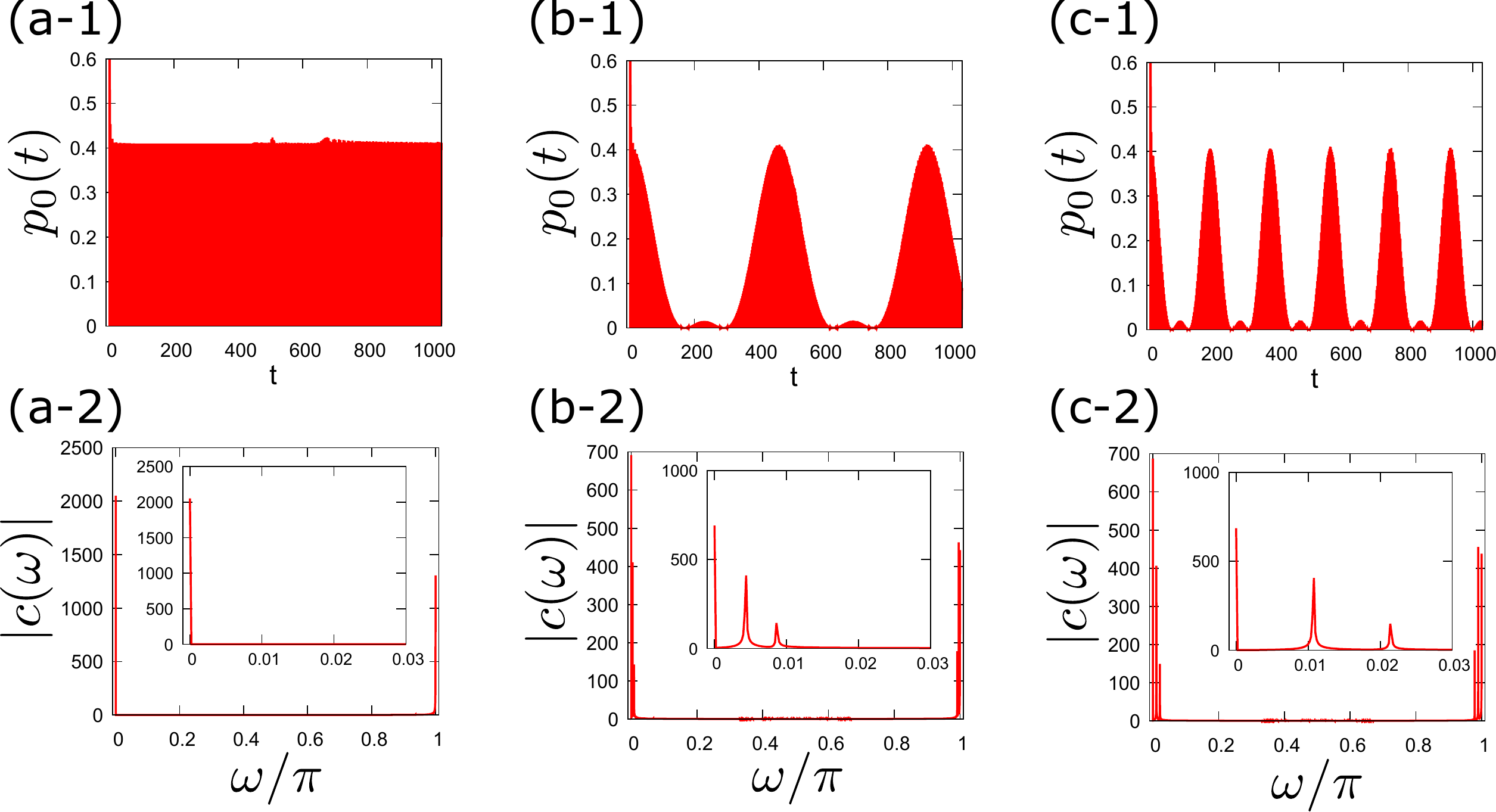}
\caption{(Top row) Time evolution of the probability \(p_0(t)\) in the case of \(\Delta\nu=3\). (Bottom row) Absolute values of the Fourier coefficient \(|c(\omega)|\) calculated from \(p_0(t)\). The insets in the right column illustrate the enlargement of the main panel near \(\frac{\omega}{\pi}=0\). (a) \(\delta=0 \ (\varepsilon_m=0.150\pi)\). (b) \(\delta=0.02 \ (\varepsilon_m=0.144\pi)\). (c) \(\delta=0.05 \ (\varepsilon_m=0,134\pi)\).}
\label{fig:prob3_lg}
\end{figure*}

We numerically calculate the probability \(p_0(t)\) by simulating the time evolution at \(\delta\ne0\).
In this section the initial state and the coin parameters in \(x\leq0\) are given as \(\ket{\psi(0)}=\ket{x=0}\otimes\frac{1}{\sqrt{2}}(\ket{L}+i\ket{R})\) and \((\theta_1^L,\theta_2^L)=(\frac{3}{4}\pi,\frac{1}{20}\pi)\), respectively.
The topological number in \(x\leq0\) is \(\nu_L=0\).
First, we fix the coin parameters for \(x>0\) as \(\theta_1^R=-\frac{1}{3}\pi, \theta_2^R=0\) (the topological number in \(x>0\) is \(\nu_R=3\)).
The time evolution of the probability \(p_0(t)\) is illustrated in the left column of Fig.\ \ref{fig:prob3_lg}.
Figure \ref{fig:prob3_lg} shows that oscillations with relatively long periods appear in the case of \(\delta\ne0\).
To clarify the oscillating nature we perform a discrete Fourier transform as
\begin{equation}
c(\omega)=\sum_{t=0}^Tp_0(t)e^{-i\omega t},~\omega=\frac{2\pi}{T+1}n,~n=0,1,...,T+1 ~,\label{eq:dft}
\end{equation}
where \(c(\omega)\) represents the complex amplitude of the mode
\(\omega\) and \(T\) denotes the total number of time steps.
In this section we set \(T=10000\).
The value of \(p_0(t)\) must be zero if \(t\) is odd because the time-evolution operator \(U_{\delta}\) contains three shift operators; then we have a mode with \(\omega=\pi\) in all cases.
In the insets of the right column of Fig.\ \ref{fig:prob3_lg}, we clearly see two extra oscillating modes near \(\omega=0\) if \(\delta\ne0\).
Similarly, we also have two extra oscillating modes near \(\omega=\pi\) if \(\delta\ne0\).
We note that the oscillating period become short as \(\delta\) increases.
These results agree well with the analysis in Appendix \ref{app:count}.
Then we conclude that these oscillating modes originate from the defective states and have the frequency \(\omega_{\delta},2\omega_{\delta},\pi-2\omega_{\delta}\), and \(\pi-\omega_{\delta}\), respectively.

\begin{figure}[tb]
\centering
\includegraphics[width=\columnwidth]{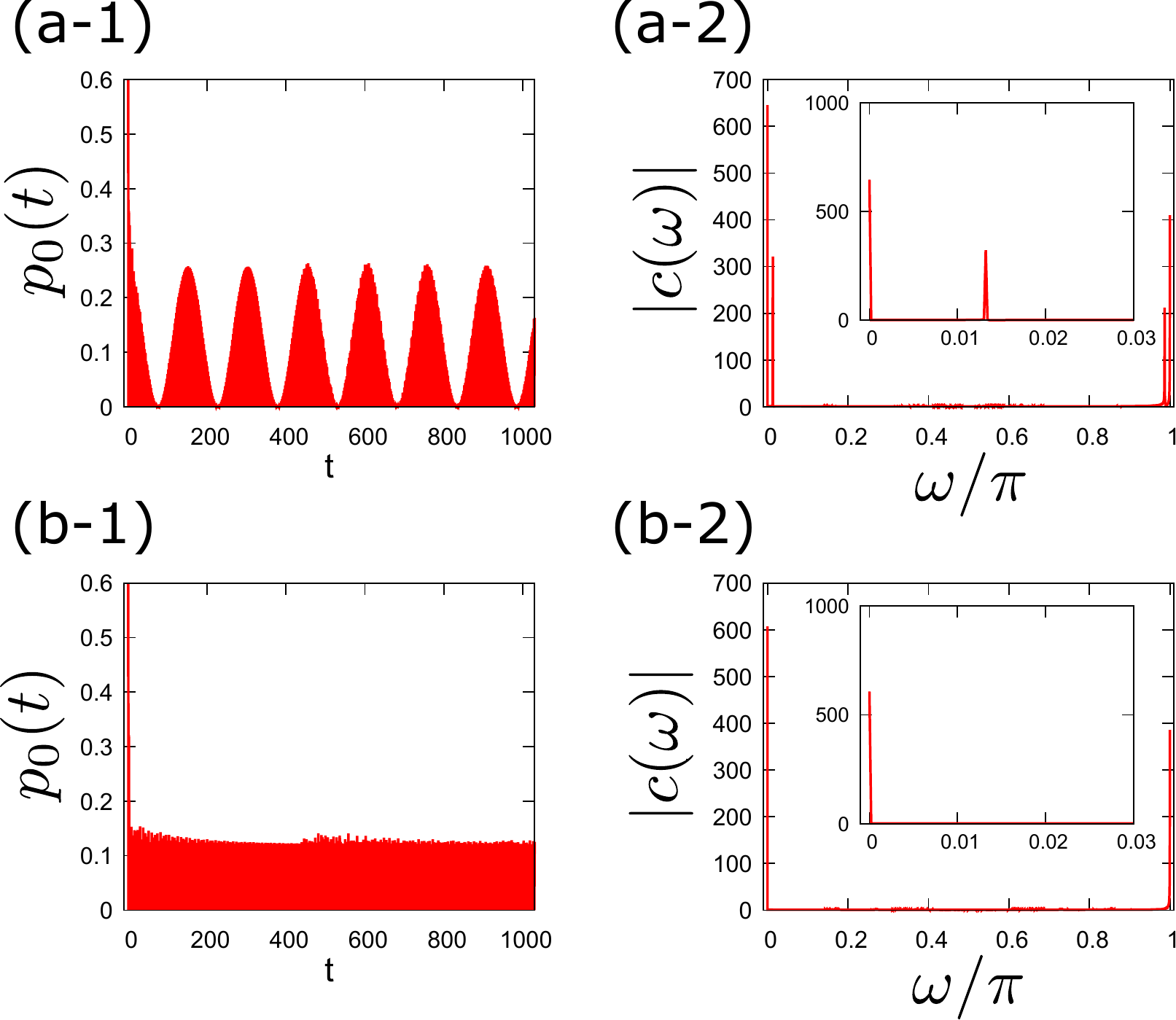}
\caption{(Left column) Time evolution of the probability \(p_0(t)\) with \(\delta=0.05\). (Right column) Absolute values of the Fourier coefficient \(|c(\omega)|\) calculated from \(p_0(t)\). (a) \(\theta_1^R=-\frac{1}{10}\pi, \theta_2^R=\frac{2}{5}\pi\) \((\Delta\nu=2, \varepsilon_m=0.134\pi)\). (b) \(\theta_1^R=-\frac{1}{15}\pi, \theta_2^R=\frac{2}{3}\pi\) \((\Delta\nu=1, \varepsilon_m=0.134\pi)\).}
\label{fig:prob12_lg}
\end{figure}

Next, we consider the cases of \(\Delta\nu=1\) and \(2\).
We calculate \(p_0(t)\) and \(c(\omega)\) in a similar way and show the results in Fig.\ \ref{fig:prob12_lg}.
Figure \ref{fig:prob12_lg} (a) shows that \(p_0(t)\) oscillates and the period is shorter than the results in Fig. \ref{fig:prob3_lg} when \(\Delta\nu=2\).
As explained in Appendix \ref{app:count}, the absence of edge states causes this difference.
Figure \ref{fig:prob12_lg} (b) shows that \(p_0(t)\) does not have extra oscillating modes unlike previous results when \(\Delta\nu=1\).
This is because there are no defective states.
Thus we can determine whether the number of edge states is one or more from the time evolution of the probability \(p_0(t)\).
We note that impurity states localized near the interface emerge in some coin parameters if \(\Delta\nu=1\).
In such cases \(p_0(t)\) oscillates due to interference between the edge states and the impurity states, and the oscillating frequency is much larger than the frequency caused by the perturbation \(\delta\).

\begin{figure*}[tb]
\centering
\includegraphics[width=2\columnwidth]{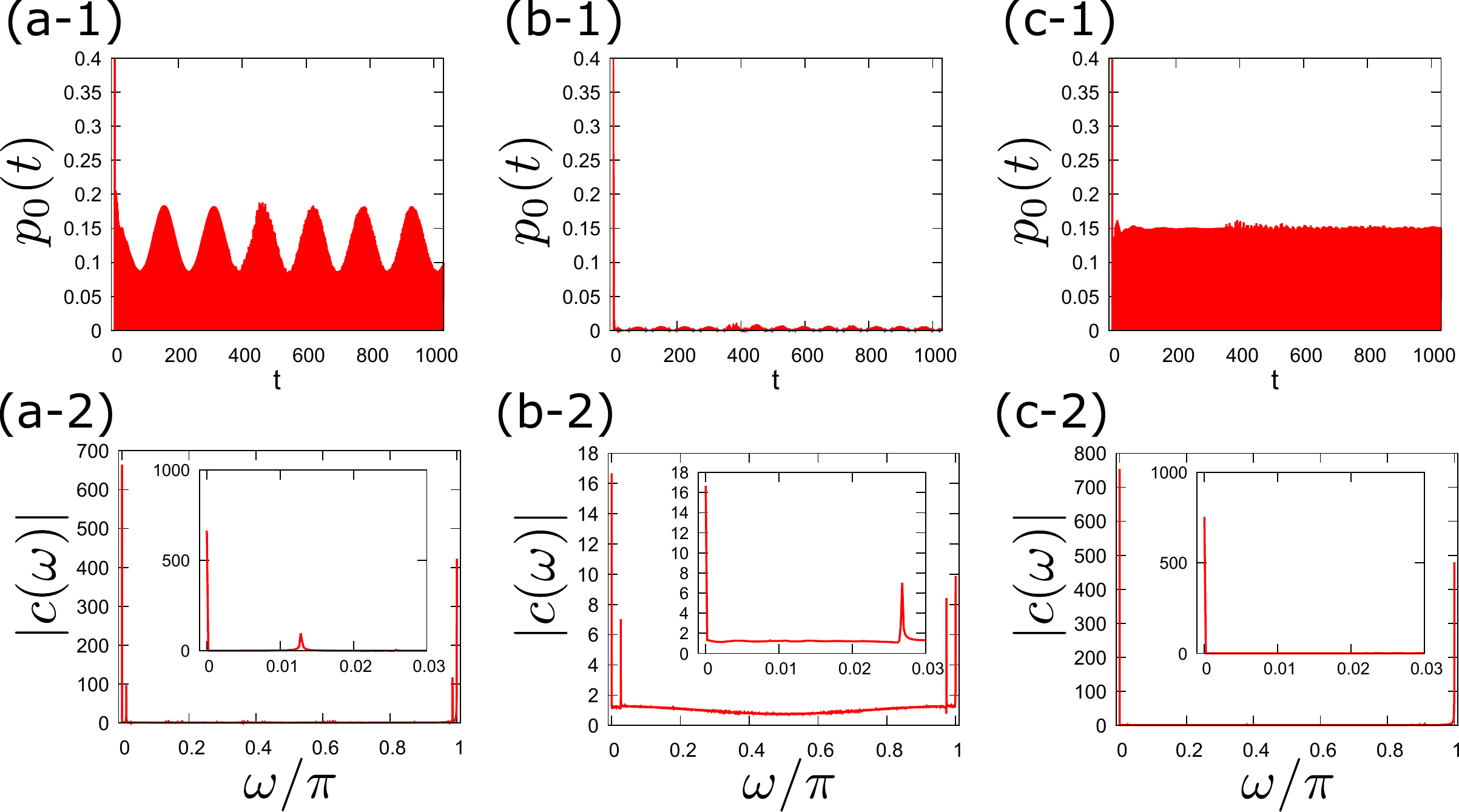}
\caption{(Top row) Time evolution of the probability \(p_0(t)\) in the case of small band gaps with \(\delta=0.05\). (Bottom row) Absolute values of the Fourier coefficient \(|c(\omega)|\) calculated from \(p_0(t)\). In each case, we set the coin parameters of the left regions as \(\theta_1^L=\frac{1}{8}\pi, \theta_2^L=\frac{1}{10}\pi\) and the initial state as \(\ket{x=0}\otimes\frac{1}{\sqrt{2}}(\ket{L}+i\ket{R})\). (a) \(\theta_1^R=-\frac{1}{5}\pi, \theta_2^R=-\frac{1}{12}\pi\) \((\Delta\nu=3, \varepsilon_m=0.0237\pi)\). (b) \(\theta_1^R=-\frac{1}{10}\pi, \theta_2^R=\frac{2}{5}\pi\) \((\Delta\nu=2, \varepsilon_m=0.0239\pi)\). (c) \(\theta_1^R=-\frac{1}{20}\pi, \theta_2^R=-\frac{1}{7}\pi\) \((\Delta\nu=1, \varepsilon_m=0.0226\pi)\).}
\label{fig:prob_sg}
\end{figure*}

In these cases, the defective states and edge states have similar properties such as localization lengths.
However, the perturbation alters the properties of the defective states if the band gaps are small.
We recalculate \(p_0(t)\) and \(c(\omega)\) in the case of small band gaps and show the results in Fig.\ \ref{fig:prob_sg}.
In these cases \(\varepsilon_m\) are very small compared to the cases of Figs.\ \ref{fig:prob3_lg} and \ref{fig:prob12_lg}.
Compared to the cases of large band gaps, the localization lengths of the defective states are much shorter than those of the edge states.
In such cases, the probability \(p_0(t)\) takes a large value and oscillates if \(\Delta\nu=3\), \(p_0(t)\) oscillates if \(\Delta\nu=2\), and \(p_0(t)\) takes a large value if \(\Delta\nu=1\).
We provide a detailed discussion in Appendix \ref{app:count}.
As shown in Fig.\ \ref{fig:prob_sg}, the time dependence of \(p_0(t)\) differs depending on \(\Delta\nu\) and these results agree well with our analysis.
Therefore, we can determine the number of edge states from time evolution of the probability \(p_0(t)\) in the cases of small band gaps.

\begin{figure*}[tb]
\centering
\includegraphics[width=2\columnwidth]{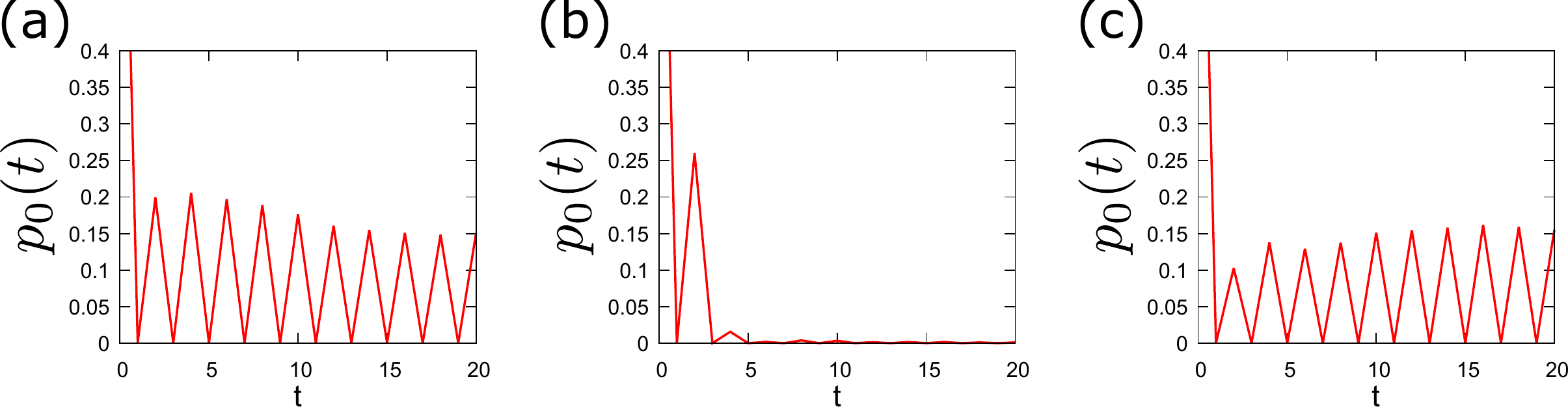}
\caption{The probability \(p_0(t)\) in shorter time steps \(t\). (a) \(\theta_1^R=-\frac{1}{5}\pi, \theta_2^R=-\frac{1}{12}\pi\) \((\Delta\nu=3)\). (b) \(\theta_1^R=-\frac{1}{10}\pi, \theta_2^R=\frac{2}{5}\pi\) \((\Delta\nu=2)\). (c) \(\theta_1^R=-\frac{1}{20}\pi, \theta_2^R=-\frac{1}{7}\pi\) \((\Delta\nu=1)\).}
\label{fig:os_exp}
\end{figure*}

These results tell us that we need at least several hundred steps to observe the oscillation of \(p_0(t)\).
Unfortunately, implementing such a long time step is not easy in the current experiments.
Instead, we focus on the value of \(p_0(t)\) at a small time step \(t\), which can be accessible in current experiments, and show the results in Fig.\ \ref{fig:os_exp}.
As shown in Fig.\ \ref{fig:os_exp}, we observe that the value of
\(p_0(t)\) is still large after several time steps if \(\Delta\nu\) is odd.
On the other hand, \(p_0(t)\) quickly decays if \(\Delta\nu\) is even.
Thus we can determine whether the number of edge states is odd or even for several time steps.
In practice, we can identify the two edge states (due to \(\Delta\nu=2\)) by the following procedure.
First, we prepare the three-step quantum walk with \(\gamma\ne0\) and \(\delta=0\), and observe a probability distribution in which a peak appears near the boundary as shown in Fig.\ \ref{fig:prob_dist} (a).
Next, by changing \(\gamma=0\) and \(\delta\ne0\), but keeping the coin parameters unchanged, we observe the probability near the boundary.
If the peak of the probability distribution near the boundary decays quickly, we can see that the number of edge states appearing at the boundary is two.

\section{Summary}
\label{sec:summary}

We have investigated the bulk-edge correspondence of $\mathcal{PT}$-symmetric open quantum systems with large topological numbers.
To this end, we have defined the non-unitary three-step quantum walk with $\mathcal{PT}$ symmetry
and other symmetries required for the BDI$^\dagger$ class in Ref.\ \cite{Kawabata_2018}.
We have numerically confirmed the validity of the bulk-edge
correspondence for this model since the number of eigenvalues
corresponding to multiple edge states agrees well with the difference in topological numbers.

We have also shown that the multiple edge states in non-unitary quantum
walks are robust against the symmetry-breaking perturbation, which
results in the $\mathbb{Z}_2$ topological phase in the class
D$^\dagger$.
This stability enhancement is unique to non-Hermitian systems with
multiple edge states and anti-linear symmetry.
Therefore, this observation establishes a new kind of breakdown of
the bulk-edge corresponded in non-Hermitian systems.
Finally, we have developed a procedure to distinguish the number of edge states from the time dependences of the probability distribution, available in standard experiments on quantum walks.
Our study contributes to the future experimental
verification of the bulk-edge correspondence in non-Hermitian systems.

As we have demonstrated, the quantum walk enables us to define non-unitary
time-evolution operators with various symmetries. Further, these
non-unitary quantum walks can be experimentally realized in the quantum
optical system \cite{Xiao_2017}. Since symmetry classes increase to 38
for topological phases in non-Hermitian systems,
studying non-unitary quantum walks with different symmetries is
important for further development of topological phases in open quantum
systems. We believe that the quantum walk will become an important arena in which to investigate fruitful novel phenomena in open quantum systems.

\section*{Acknowledgements}

We thank Y.\ Asano, T.\ Bessho, H.\ Hirori, R.\ Okamoto, and K.\ Yakubo for helpful discussions.
This work was supported by KAKENHI (Grants
No.\ JP18J20727, No.\ JP19H01838, No.\ JP18H01140, No.\ JP18K18733, and
No.\ JP19K03646) and a Grant-in-Aid for Scientific Research on Innovative Areas (KAKENHI Grant
No.\ JP15H05855 and No.\ JP18H04210) from the Japan Society for the Promotion of Science.

\appendix
\section{Detailed calculations of perturbed probabilities}
\label{app:count}
In this appendix, we derive the time dependence of the probability at
the interface, where edge states dominate the dynamics, when the unitary perturbed time-evolution operator $U_\delta$ in Eq.\ (\ref{eq:u_delta}) is employed.
We first consider the case of \(\Delta\nu=3\) where three
edge states appear at the interface at the same quasi-energy when $\delta=0$. The quasi-energies of defective states are \(\pm\omega_{\delta}\) or \(\pi\pm\omega_{\delta}\) because of particle-hole symmetry. We write the states whose eigenenergy is \(\omega\) or \(\omega\pm\omega_{\delta}\) as \(\ket{\psi_\omega},\ket{\psi_{\omega\pm}}\), where \(\omega=0,\pi\). We prepare an initial state \(\ket{\psi(t=0)}\) that localizes at \(x=0\), then we can expand \(\ket{\psi(t=0)}\) as a linear combination of the eigenstates of the time-evolution operator in Eq.\ \eqref{eq:u_delta},
\begin{equation}
\begin{split}
\ket{\psi(t=0)}=&a\ket{\psi_0}+b\ket{\psi_{0+}}+c\ket{\psi_{0-}}\\&+p\ket{\psi_{\pi}}
+q\ket{\psi_{\pi+}}+r\ket{\psi_{\pi-}}\\&+(\mbox{linear combination of bulk states}),
\end{split}
\end{equation}
where $a,b,c,p,q,r$ denote wavefunction amplitudes.
The state at time step \(t\) is written as
\begin{equation}
\begin{split}
\ket{\psi(t)}=&a\ket{\psi_0}+be^{-i\omega_{\delta} t}\ket{\psi_{0+}}+ce^{i\omega_{\delta} t}\ket{\psi_{0-}}+p(-1)^t\ket{\psi_{\pi}}
\\&+qe^{-i(\pi+\omega_{\delta})
 t}\ket{\psi_{\pi+}}+re^{-i(\pi -\omega_{\delta}) t}\ket{\psi_{\pi-}}\\ &+(\mbox{linear combination of bulk states}). \label{eq:psit}
\end{split}
\end{equation}
Then, the wavefunction at the interface (\(x=0\)) is given by:
\begin{equation}
\begin{split}
\psi(t)=&a\psi_0+be^{-i\omega_{\delta} t}\psi_{0+}+ce^{i\omega_{\delta} t}\psi_{0-}+p(-1)^t\psi_{\pi}
\\&+qe^{-i(\pi+\omega_{\delta})
 t}\psi_{\pi+}+re^{-i(\pi-\omega_{\delta}) t}\psi_{\pi-}\\ &+(\mbox{linear combination of bulk wavefunctions}),\label{eq:wf}
\end{split}
\end{equation}
where \(\psi(t)=\braket{x=0|\psi(t)}\) and \(\psi_j=\braket{x=0|\psi_j}\).
Edge states and defective states are localized at \(x=0\) and have
large amplitudes of wavefunctions.
On the other hand, probabilities for bulk states at \(x=0\) rapidly
decrease and we neglect contributions from bulk wavefunctions in
Eq.\ \eqref{eq:wf}.
Then, the probability of the walker at \(x=0\), \(p_0(t)=|\psi(t)|^2\) has oscillating terms with frequencies \(\omega_{\delta},2\omega_{\delta},\pi-2\omega_{\delta},\pi-\omega_{\delta}\), and \(\pi\).
We can calculate \(p_0(t)\) in the case of \(\Delta\nu=2\) similarly. We
expand the initial state localized at \(x=0\) as
\begin{equation}
\begin{split}
\ket{\psi(t=0)}=&a'\ket{\psi_{0+}}+b'\ket{\psi_{0-}}+p'\ket{\psi_{\pi+}}+q'\ket{\psi_{\pi-}} \\
&+(\mbox{linear combination of bulk states}), \label{eq:state_nu2}
\end{split}
\end{equation}
where $a', b', p', q'$ denote wavefunction amplitudes.
We define the quasi-energies of the defective states \(\pm\omega_{\delta}\) or \(\pi\pm\omega_{\delta}\) again.
Then we can see that the probability \(p_0(t)\) has oscillating terms with frequencies \(2\omega_{\delta},\pi-2\omega_{\delta}\), and \(\pi\).
Oscillating terms with frequencies \(\omega_{\delta}\) and \(\pi-\omega_{\delta}\) do not appear because there is no edge state in the case of \(\Delta\nu=2\).
In the case of \(\Delta\nu=1\), the probability \(p_0(t)\) does not
oscillate as with \(\Delta\nu=2,3\) because of there being no defective states.

\begin{figure}[tb]
\centering
\includegraphics[width=0.5\columnwidth]{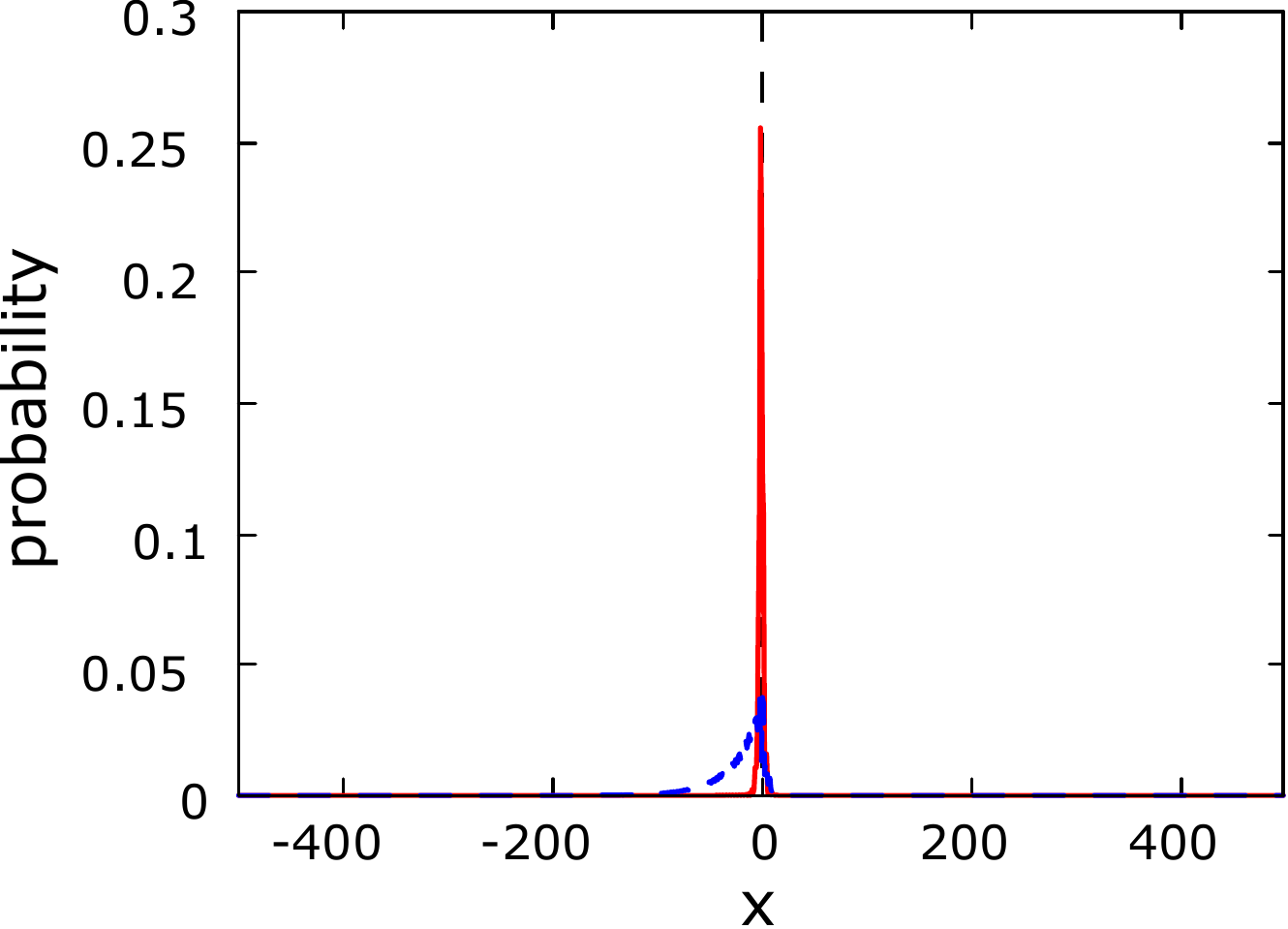}
\caption{Spatial distributions of squared absolute values of
 wavefunction amplitudes of the edge state \(\psi_0(x)\) (red solid line) and the defective state \(\psi_{0-}(x)\) (blue dashed line) in the case of small band gaps. We set the parameters \(\theta_1^L=\frac{1}{8}\pi, \theta_2^L=\frac{1}{10}\pi, \theta_1^R=-\frac{1}{5}\pi\), and \(\theta_2^R=-\frac{1}{12}\pi\).}
\label{fig:defect}
\end{figure}

If the band gap is small, the defective state has a broad localization length compared to edge states as shown in Fig.\ \ref{fig:defect}.
In such cases, the wavefunctions at the interface satisfy \(|\psi_{0}|,|\psi_{\pi}|\gg|\psi_{0\pm}|,|\psi_{\pi\pm}|\).
Then the relation \(|\psi_{j}^*\psi_{j'}|\gg|\psi_{j}^*\psi_{j\pm}|\gg|\psi_{j\pm}^*\psi_{j'\pm}|\) is also satisfied, where \(j,j'\) takes \(0\) or \(\pi\).
These terms constitute constant terms, oscillating terms with frequency \(\omega_{\delta}\) and \(\pi-\omega_{\delta}\) and oscillating terms with frequency \(2\omega_{\delta}\) and \(\pi-2\omega_{\delta}\), respectively.
From this relation we observe that oscillating terms with frequency \(2\omega_{\delta}\) or \(\pi-2\omega_{\delta}\) are negligible if edge states emerge.
We also note that a pair of defective states is necessary to make \(p_0(t)\) oscillate.
Summarizing this discussion, we expect that \(p_0(t)\) takes a large value if there are edge states and oscillates if there are defective states.

\end{document}